\documentclass{aastex63p}
\newcommand{\ms}{\mbox{m\ \ensuremath{\rm{s}^{-1}}}}

\accepted{2020 October 29th}
\shorttitle{Dynamical Instability of Titan's Thermospheric Jet}

\begin{document}

\title{Detection of Dynamical Instability in Titan's Thermospheric Jet}

\correspondingauthor{M. A. Cordiner}
\email{martin.cordiner@nasa.gov}

\author{M. A. Cordiner}
\affiliation{Solar System Exploration Division, NASA Goddard Space Flight Center, 8800 Greenbelt Road, Greenbelt, MD 20771, USA.}
\affiliation{Department of Physics, Catholic University of America, Washington, DC 20064, USA.}

\author{E. Garcia-Berrios}
\affiliation{Solar System Exploration Division, NASA Goddard Space Flight Center, 8800 Greenbelt Road, Greenbelt, MD 20771, USA.}
\affiliation{Department of Physics, Catholic University of America, Washington, DC 20064, USA.}

\author{R. G. Cosentino}
\affiliation{Solar System Exploration Division, NASA Goddard Space Flight Center, 8800 Greenbelt Road, Greenbelt, MD 20771, USA.}
\affiliation{Department of Astronomy, University of Maryland, College Park, MD 20742, USA.}

\author{N. A. Teanby}
\affiliation{School of Earth Sciences, University of Bristol, Wills Memorial Building, Queens Road, Bristol, BS8 1RJ, UK.}

\author{C. E. Newman}
\affiliation{Aeolis Research, 333 N Dobson Road, Unit 5, Chandler, AZ 85224, USA.}

\author{C. A. Nixon}
\affiliation{Solar System Exploration Division, NASA Goddard Space Flight Center, 8800 Greenbelt Road, Greenbelt, MD 20771, USA.}

\author{A. E. Thelen}
\affiliation{Solar System Exploration Division, NASA Goddard Space Flight Center, 8800 Greenbelt Road, Greenbelt, MD 20771, USA.}

\author{S. B. Charnley}
\affiliation{Solar System Exploration Division, NASA Goddard Space Flight Center, 8800 Greenbelt Road, Greenbelt, MD 20771, USA.}



\begin{abstract}

Similar to Earth, Saturn's largest moon, Titan, possesses a system of high-altitude zonal winds (or jets) that encircle the globe. Using the Atacama Large Millimeter/submillimeter Array (ALMA) in August 2016, \citet{lel19} discovered an equatorial jet at much higher altitudes than previously known, with a surprisingly fast speed of up to $\sim340$~\ms, but the origin of such high velocities is not yet understood. We obtained spectrally and spatially resolved ALMA observations in May 2017 to map Titan's 3D global wind field and compare our results with a reanalysis of the August 2016 data. Doppler wind velocity maps were derived in the altitude range $\sim~300$--1000~km (from the upper stratosphere to the thermosphere).  At the highest, thermospheric altitudes, a 47\% reduction in the equatorial zonal wind speed was measured over the 9-month period (corresponding to $L_s=82^{\circ}$--$90^{\circ}$ on Titan). This is interpreted as due to a dramatic slowing and loss of confinement (broadening) of the recently-discovered thermospheric equatorial jet, as a result of dynamical instability. {These unexpectedly-rapid changes in the upper-atmospheric dynamics are consistent with strong variability of the jet's primary driving mechanism}.

\end{abstract}

\keywords{planets and satellites: atmospheres --- planets and satellites: individual (Titan) --- techniques: imaging spectroscopy --- submillimeter: planetary systems}


\section{Introduction}

Titan is noteworthy in our Solar System due to its unusually thick, dense (1.45 bar) atmosphere comprised predominantly of molecular nitrogen and methane, subject to a complex atmospheric chemistry driven by solar radiation, cosmic rays, and charged particles from Saturn's magnetosphere. Titan also has a relatively slow rotation rate (15.9 Earth days), intermediate between that of Earth/Mars ($\approx1$ day) and the very slow rotation rate of Venus (243 Earth days), which like Titan displays strong atmospheric superrotation. Studies of Titan can therefore provide unique insights into fundamental atmospheric processes such as photochemistry, winds and global circulation in a unique physical and chemical environment, {more analogous to Earth than the other terrestrial planets} \citep{hor17}.

The presence of prograde (superrotating) zonal winds in Titan's troposphere/stratosphere was initially inferred from stellar occultation observations \citep{hub93}. This was confirmed by in-situ Doppler measurements during the descent of the Cassini-Huygens probe \citep{bir05}, which detected eastward winds with speeds up to 100~\ms, for altitudes $z<150$~km. The first direct detection of zonal winds in the upper-stratosphere/lower-mesosphere ($z=300$--450~km) was by \citet{mor05}, using spatially resolved, ground-based microwave spectroscopy with the Plateau de Bure interferometer. Titan's zonal winds are responsible for the longitudinal transport/mixing of gases produced as a result of photochemistry on Titan's day-side, leading to significant east-west asymmetries for the short-lived HNC molecule \citep{cor19}, but the temporal variability of Titan's high-altitude winds (in the mesosphere and above), remains to be studied in detail.

\citet{lel19} used the Atacama Large Millimeter/submillimeter Array (ALMA) to derive {Doppler wind maps for six gases, covering eight different altitudes in the range $z\sim300$--1000~km}. The combined spectral/spatial resolution and sensitivity of { ALMA allowed} the Doppler shifts of the {observed} emission lines to be measured as a function of position across Titan's disk, from which the {zonal wind speeds as a function of latitude and altitude were derived --- altitudinal information coming from the particular vertical abundance distribution and spectral line opacity of each gas}. An unexpectedly intense, superrotating equatorial wind/jet was found, which was strongest in the thermosphere. With a velocity of $\sim340$~\ms, such a fast zonal wind defies explanation using current theoretical models for general circulation \citep{ris00,mul00,mul08}, which predict thermospheric wind speeds of only $\sim60$~\ms. 

Here we present {new} measurements of Titan's high-altitude (stratospheric-thermospheric) wind field, obtained through ALMA interferometric observations of three atmospheric gases (CH$_3$CN, HC$_3$N and HNC) in {May 2017 (corresponding to planetocentric solar longitude $L_s=90^{\circ}$; Titan's northern summer solstice), compared with a reanalysis of the \citet{lel19} data from August 2016 ($L_s=82^{\circ}$)}. This has enabled the first study of temporal variability in Titan's thermospheric jet, leading to new insights regarding its (in)stability and possible driving mechanisms.

\section{Observations}

Interferometric observations of Titan were carried out using 46 antennas of the main (12 m) array on 2017-05-08 and 2017-05-16, as part of ALMA program 2016.A.00014.S (PI: M. Cordiner). The correlator was configured to observe several Band 7 frequency windows in the range 349--364~GHz. The HNC ($J=4-3$) and HC$_3$N ($J=40-39$) transitions were observed at a spectral resolution of $\Delta{\nu}=244$~kHz (0.2~km\,s$^{-1}$), while the CH$_3$CN ($J=19-18$) band was observed at $\Delta{\nu}=488$~kHz (0.4~km\,s$^{-1}$), with two channels per spectral resolution element. The antenna configuration was moderately extended, resulting in a spatial resolution of $\approx0.24\times0.19''$ (using natural visibility weighting). The total on-source observing time was 138~min, leading to RMS noise levels $\approx2$~mJy\,beam$^{-1}$\,MHz$^{-1}$. For additional details on these observations see \citet{cor19}. 

{To ensure consistency of the data analysis methodology between 2016 and 2017, we re-analyzed the CH$_3$CN, HC$_3$N and HNC data observed by \citet{lel19} on 2016-08-18.} Raw data were obtained from the ALMA archive (project 2015.1.01023.S; PI: M. Gurwell), covering HNC ($J=4-3$, with $\Delta{\nu}=122$~kHz) and CH$_3$CN ($J=19-18$, with $\Delta{\nu}=122$~kHz and 488~kHz). These data also included the same HC$_3$N ($J=40-39$) transition observed in 2017, but at a lower spectral resolution ($\Delta{\nu}=977$~kHz, insufficient for accurate radial velocity measurements), so we instead used observations of the HC$_3$N ($J=39-38$) line at 354,697~MHz (with $\Delta{\nu}=122$~kHz), obtained on 2016-08-19 as part of the same program.

The raw observations were first continuum-subtracted using low-order polynomial fits to the line-free spectral regions adjacent to our lines of interest. The data were then imaged and cleaned (deconvolved) using the {\tt tclean} algorithm in CASA 5.6 \citep{emo19}. This routine implicitly corrects for the time-variability of Titan's position and radial velocity with respect to the observer, so the resulting image cubes are in the rest frame of Titan's center-of-mass. Reliability of the rest frequency scale was established based on the observation that the emission line Doppler shifts were typically close to zero (within the uncertainties) along the lines of sight intersecting Titan's polar axis.  The image pixel size was set to $0.025''$, with a flux threshold of twice the RMS noise per channel, and a deconvolution mask diameter of $1.3''$, encircling the entirety of the detected flux. To facilitate intercomparison of the 2016 and 2017 images, accounting for Titan's varying Geocentric distance ($\Delta$), both datasets were convolved to common circular beam dimensions ($\theta_{km}$), as projected in the plane of the sky at $\Delta=9.75$~au in 2016 and $\Delta=9.21$~au in 2017, resulting in $\theta_{km}=1670$~km for CH$_3$CN, $\theta_{km}=1570$~km for HC$_3$N and $\theta_{km}=1600$~km for HNC {(compared with Titan's 5150~km; $0.72''$--$0.77''$ diameter)}.

\subsection{Doppler Mapping}

\begin{figure}
\centering
\includegraphics[width=0.45\textwidth]{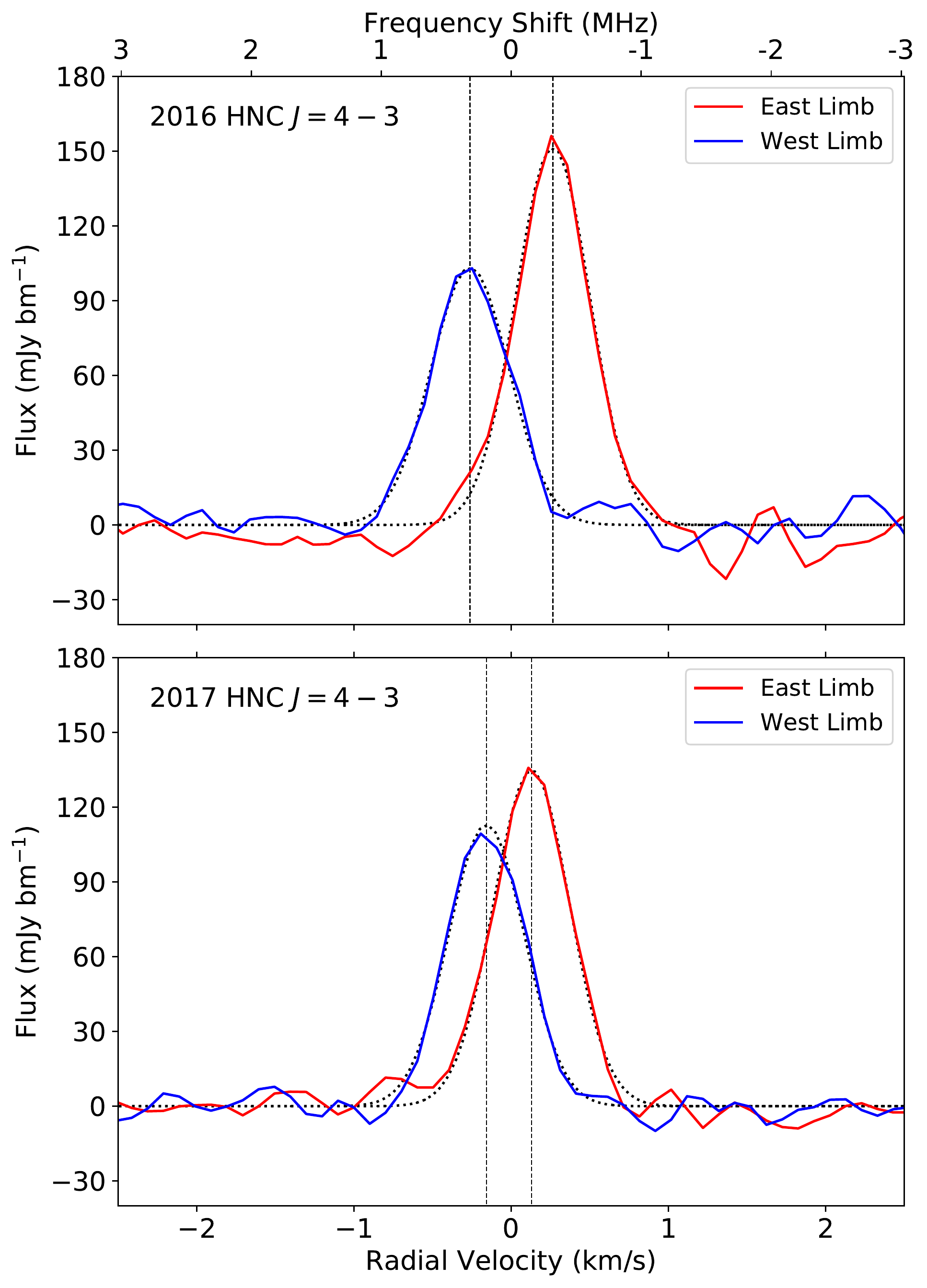}
\hspace{3mm}
\includegraphics[width=0.42\textwidth]{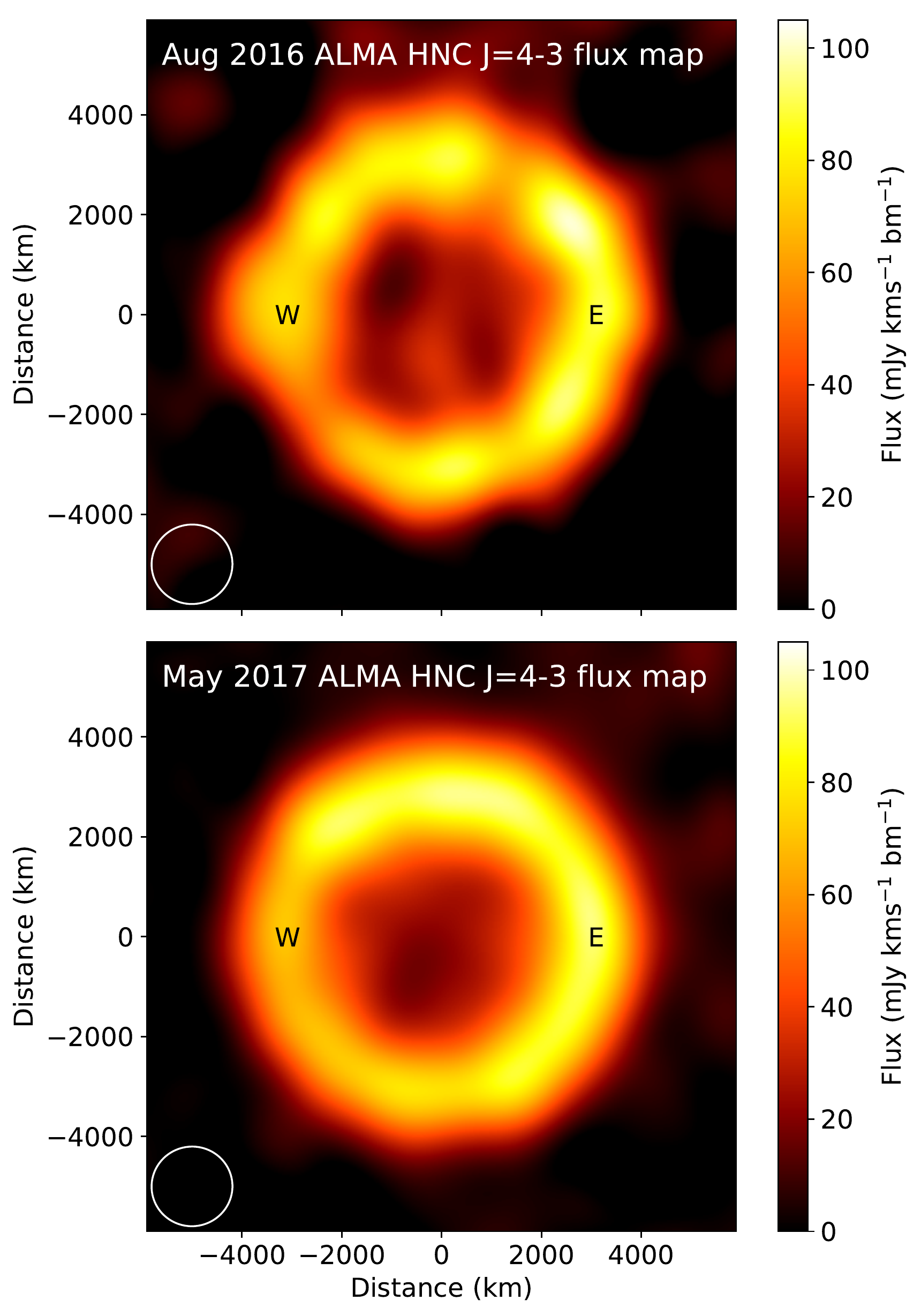}
\caption{Left panel: ALMA HNC ($J=4-3$) spectra observed at Titan's eastern and western limbs in August 2016 and May 2017, rebinned to a common channel spacing of 202 m\,s$^{-1}$. Fitted Gaussian profiles are overlaid (dotted curves), along with their central velocities (vertical dotted lines). Right panels: corresponding continuum-subtracted, spectrally-integrated flux maps of Titan's HNC ($J=4-3$) emission. The eastern and western spectra in the left panels were extracted {600 km above the equator, at the points labeled E and W, respectively}. The FWHM of the (Gaussian) spatial resolution element is shown lower left of each map. Sky-projected distances are given with respect to the center of Titan's disk (the ordinate axis is aligned with celestial north). \label{fig:hnc}}
\end{figure}

We followed the method of \citet{lel19} to determine radial velocities (along the line of sight) for each gas as a function of spatial coordinate. This involved extracting individual spectra from the 2016 and 2017 ALMA image cubes and fitting the emission lines with Gaussians to determine their Doppler shifts. Since the observed limb emission is predominantly optically thin, and originates from sufficiently high altitudes that thermal broadening dominates the line core shapes, the Gaussian Doppler shift provides a direct measure of the component of the wind velocity along the line-of-sight. A demonstration of this technique is shown for HNC in Figure \ref{fig:hnc}, resulting in a mean absolute Doppler shift of $265\pm11$~\ms\ in 2016 and $142\pm6$~\ms\ in 2017 (averaged over the east and west limbs).

As shown by \citet{cor19} and \citet{lel19}, pressure-broadened Lorentzian line wings are apparent for HC$_3$N and CH$_3$CN, so for these species, the fits were restricted to narrow (3~MHz wide) spectral regions, centered on each of the line cores. A constant (additive) intensity offset was also accounted for, to remove the pseudo-continuum created by the broad emission line wings, enabling more-accurate Gaussian fits to the narrow (FWHM $<1.4$~MHz) line cores. For CH$_3$CN, the results for the three highest-frequency lines ($J_K=19_0-18_0,\ 19_1-18_1,\ 19_2-18_2$) were averaged together for improved sensitivity.

The uncertainty on each velocity measurement was derived using {a similar Monte Carlo approach to \citet{lel19}}, re-fitting the Gaussian line model to a large set of (300) noisy, synthetic datasets. The $\pm68^{th}$ percentiles of the resulting parameter distributions are interpreted as $1\sigma$ errors. 

\section{Results}

\begin{figure}
\includegraphics[height=50mm]{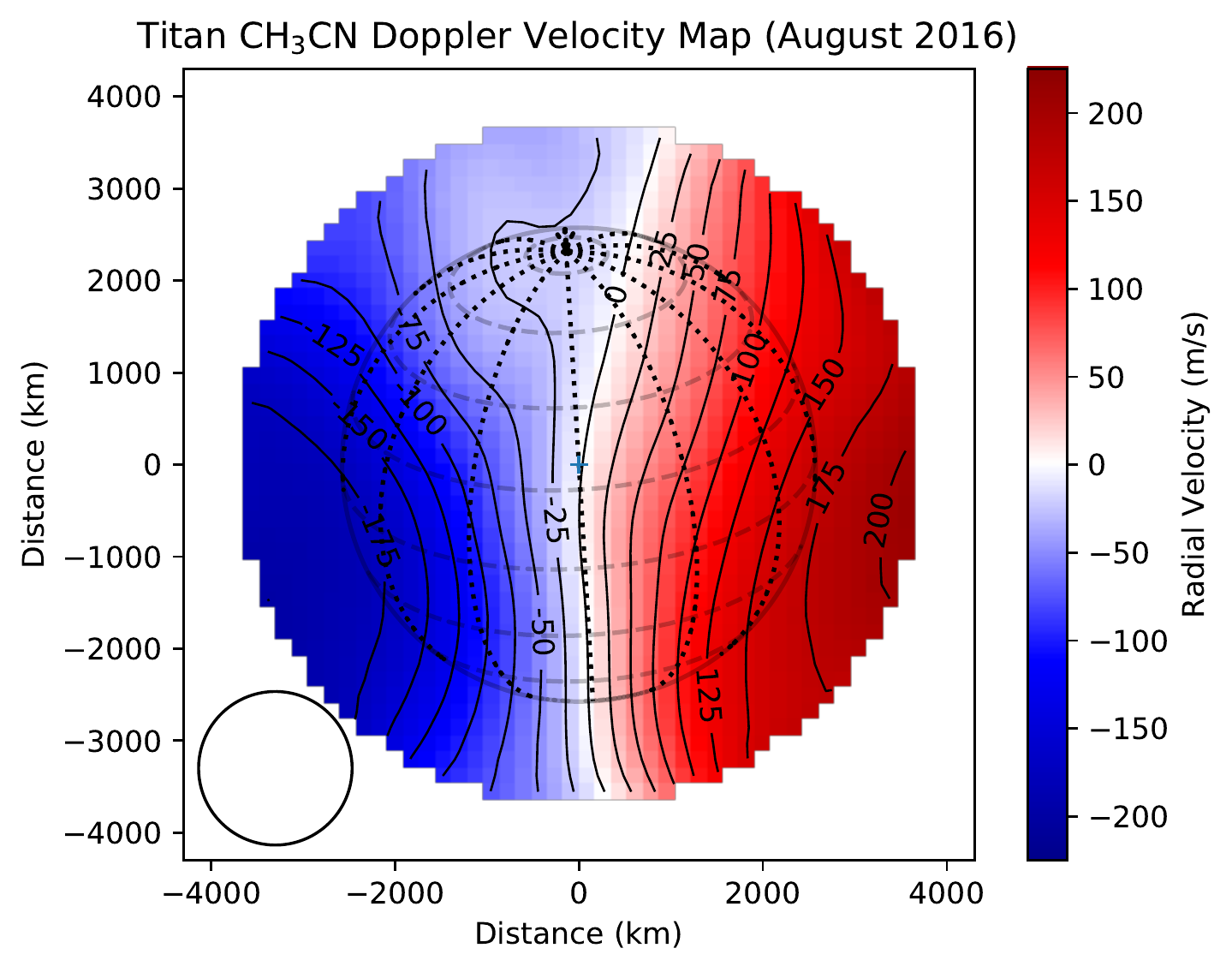}
\includegraphics[height=50mm]{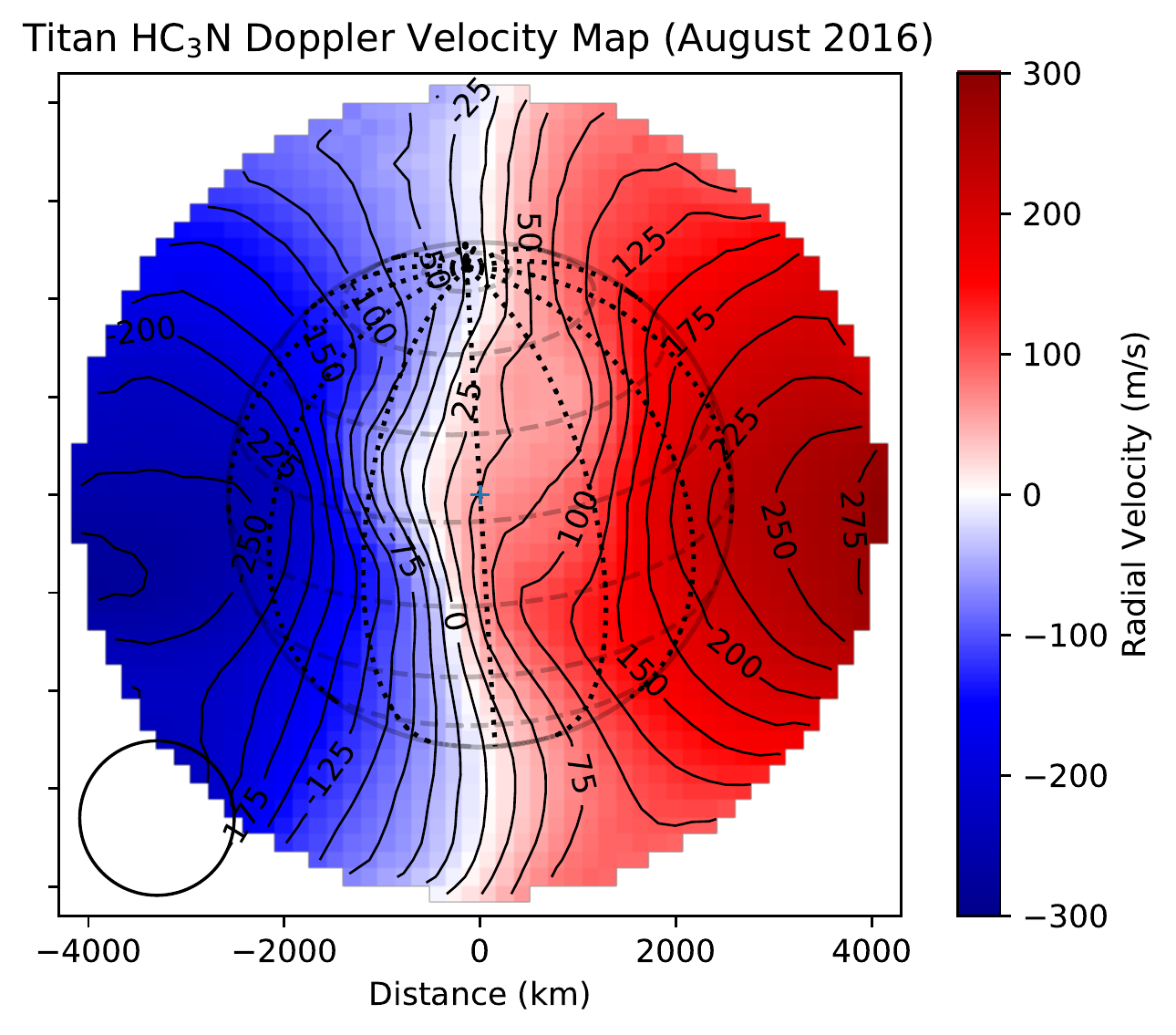}
\includegraphics[height=50mm]{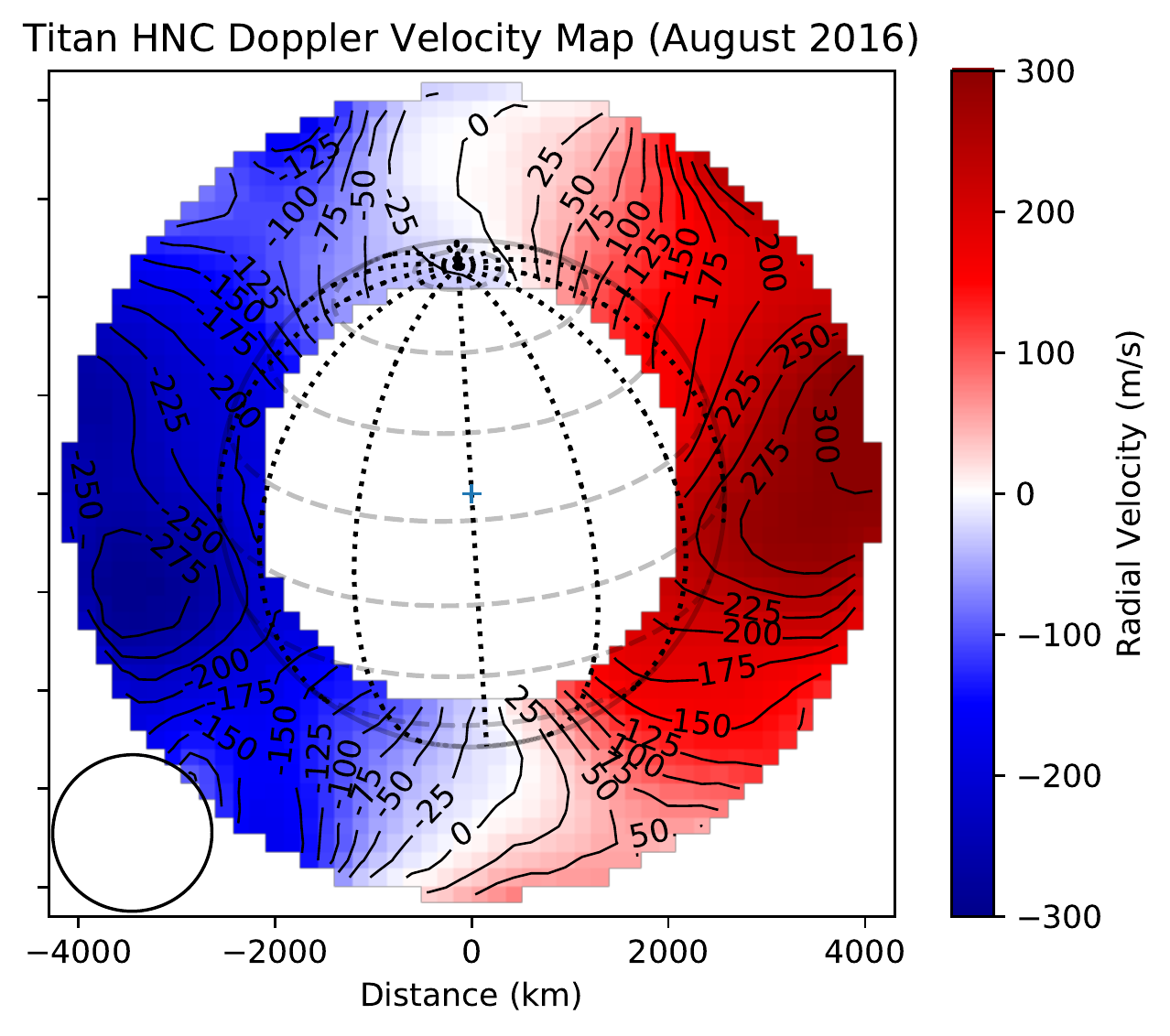}\\
\includegraphics[height=50mm]{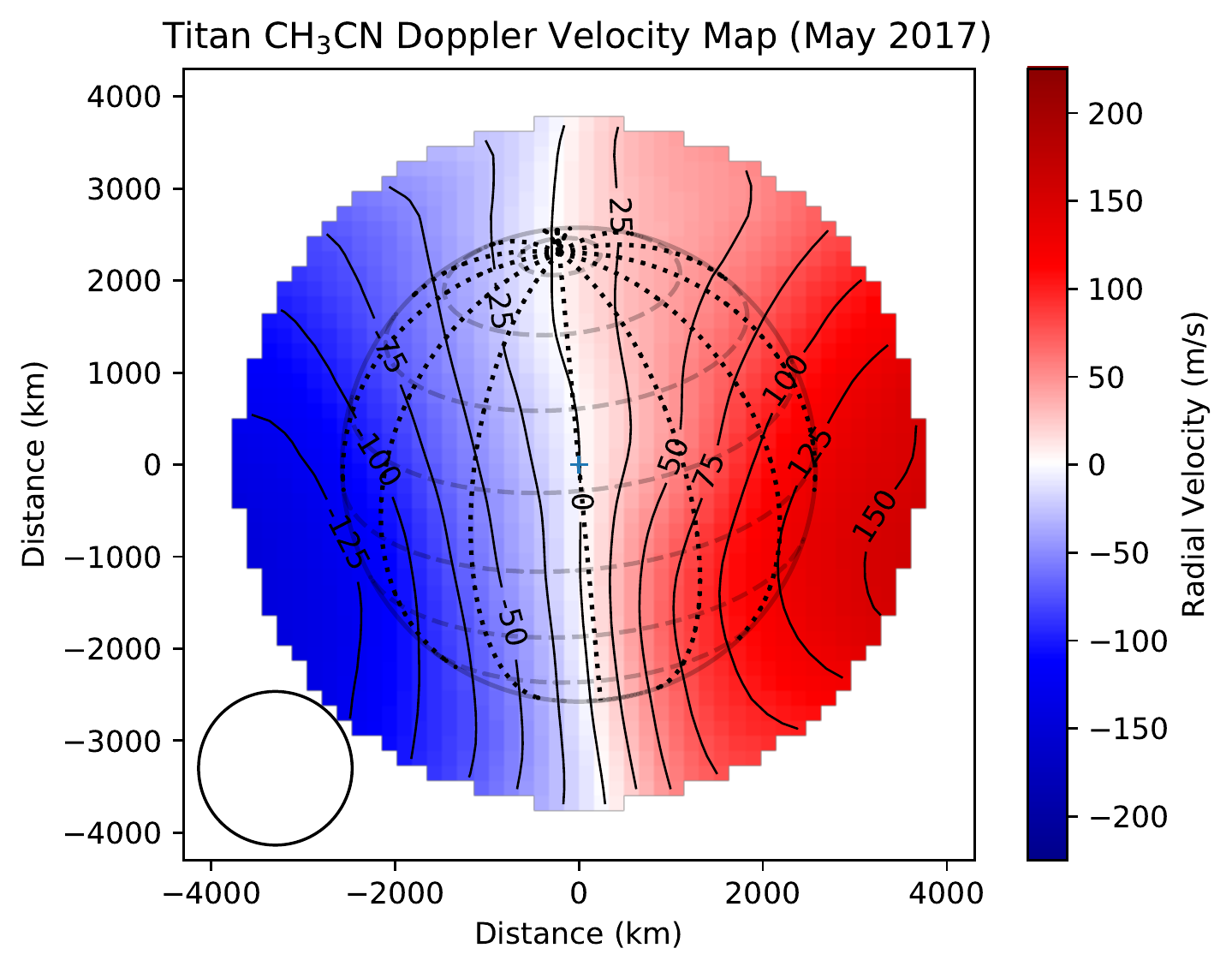}
\includegraphics[height=50mm]{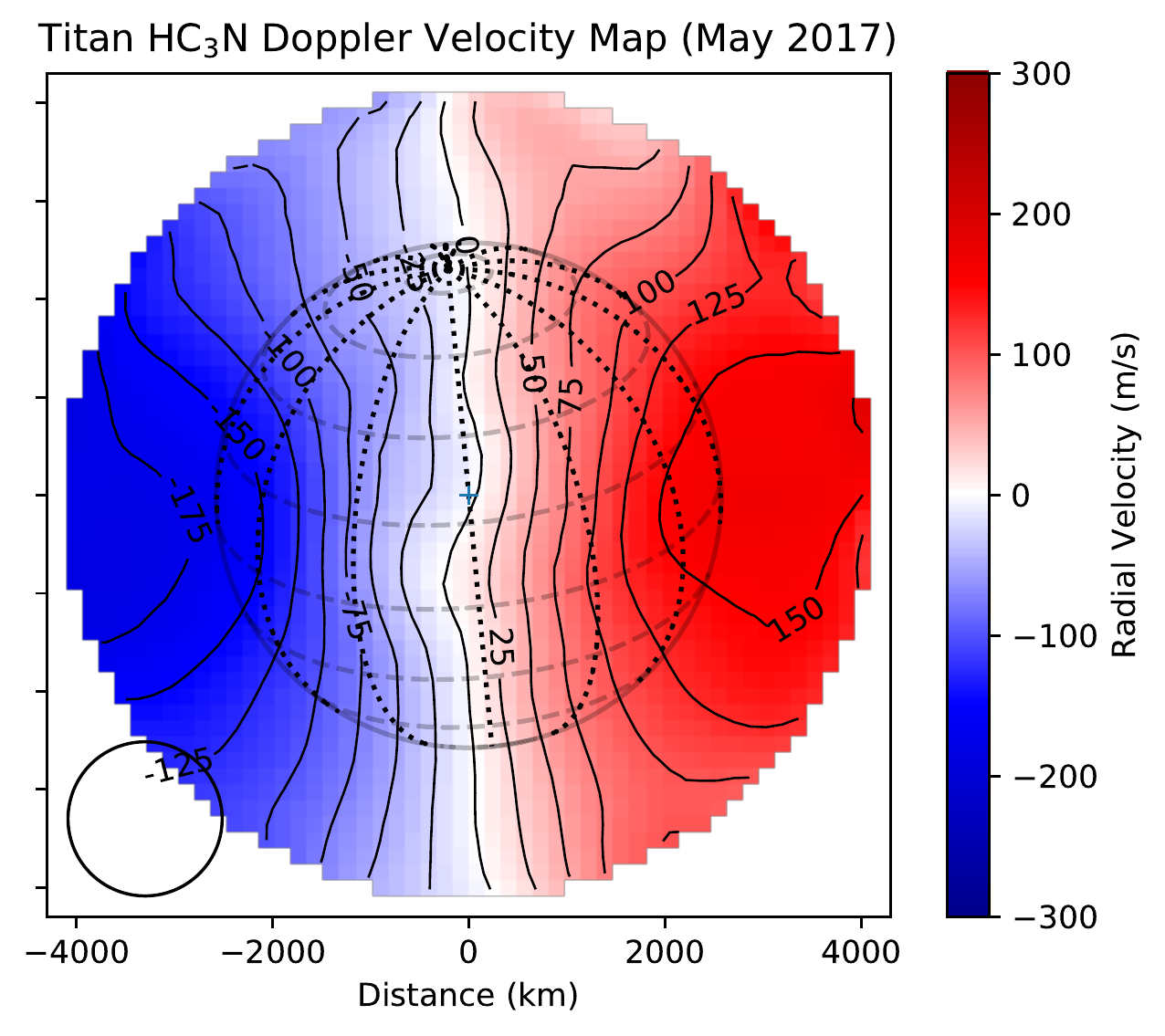}
\includegraphics[height=50mm]{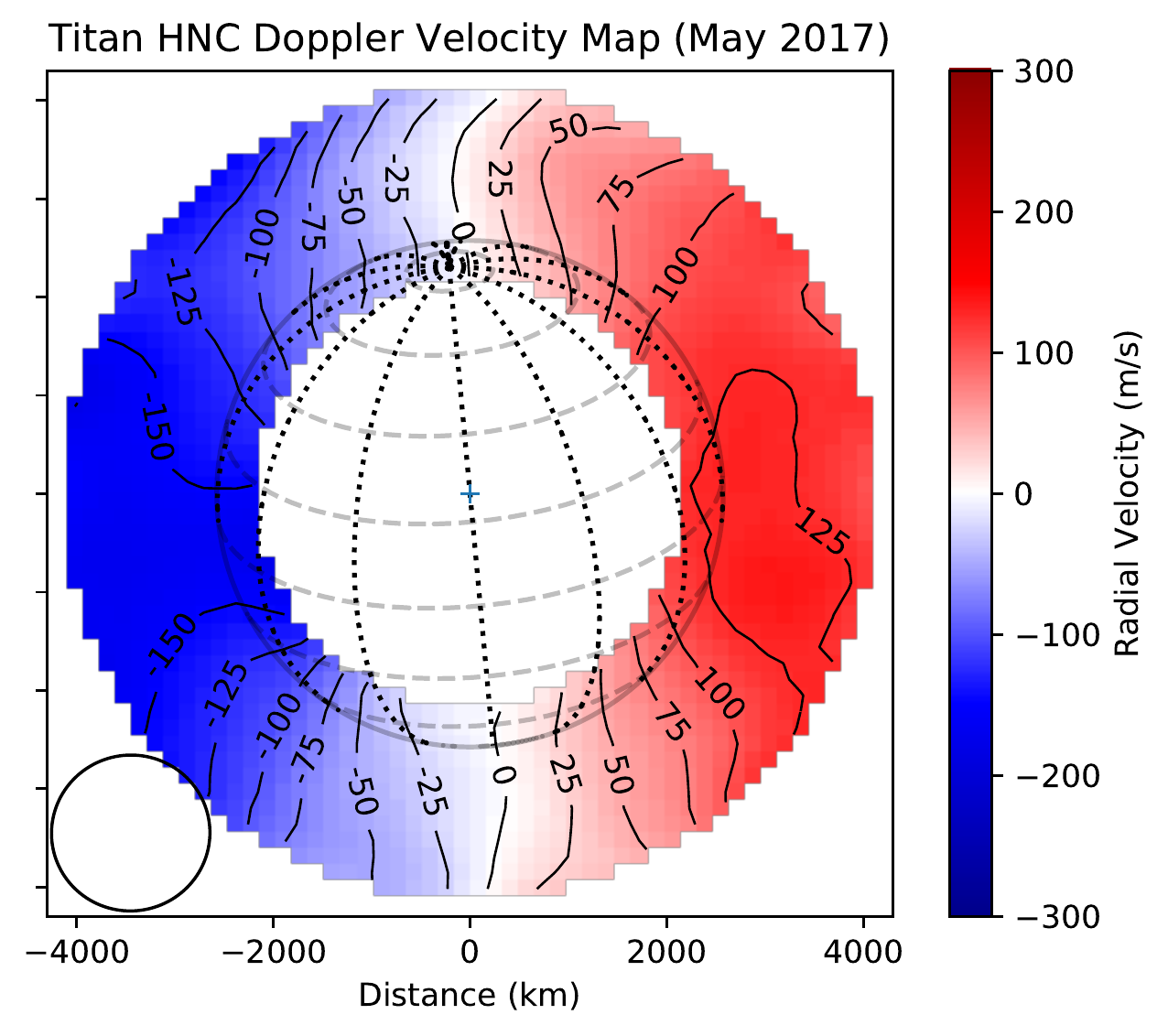}
\caption{Doppler wind velocity maps based on ALMA observations of CH$_3$CN, HC$_3$N and HNC in August 2016 ($L_s=82^{\circ}$) and May 2017 ($L_s=90^{\circ}$). Velocity contours are labeled in units of m\,s$^{-1}$. Wire-frame sphere indicates Titan's orientation with respect to the ALMA field of view. The spatial resolution (FWHM of the elliptical Gaussian beam) is shown lower-left for each panel. The data for these figures are available in FITS format from the following URL: https://doi.org/10.5281/zenodo.3965956. \label{fig:obs}}
\end{figure}

Two-dimensional Doppler maps for CH$_3$CN, HC$_3$N and HNC are shown in Figure \ref{fig:obs}, where the color of each pixel indicates the line-of-sight radial velocity with respect to Titan's rest frame. The maps have been masked to only show the circular/annular regions where {the total signal-to-noise ratio was $>10$}. The average ($\pm1\sigma$) velocity uncertainties in the masked regions are as follows: CH$_3$CN (2016): 5.7~\ms, CH$_3$CN (2017): 2.4~\ms, HC$_3$N (2016): 11.0~\ms, HC$_3$N (2017): 4.9~\ms, HNC (2016): 19.6~\ms, HNC (2017): 7.5~\ms. {Velocity error maps are given in Appendix A, Figure 5.} Additional systematic uncertainties due to errors on the molecular line rest frequencies are: 0.3~\ms\ for CH$_3$CN, 2--8~\ms\ for HC$_3$N and 7~\ms\ for HNC \citep{pic98,mul01}.

The Doppler maps are dominated by a west-to-east zonal wind pattern, {and for 2016, our results are closely consistent with those obtained by \citet{lel19}; small differences are explained as a result of convolving their data to a slightly lower resolution}. Wind velocities are highest toward lower latitudes, and the thermospheric jet is clearly evident from the tight curvature of the velocity contours about the equatorial limb region for HC$_3$N and HNC in 2016. 

Figure 3 of \citet{lel19} shows the altitudes from which the detected limb emission originated for each species: CH$_3$CN emission is from a mean altitude of $\bar{z}=345$~km, HC$_3$N from $\bar{z}=710$~km and HNC from $\bar{z}=990$~km. {The mean emission altitudes calculated by \cite{cor19} for May 2017 were very similar (using a slightly different method), which is consistent with (1) the close similarity between the equatorial vertical abundance profiles retrieved for the two epochs, and (2) the observed stability in Titan's equatorial molecular abundances over a period of several years \citep{the19,tea19}}. In 2016, however, there was a clear trend for increasing equatorial zonal wind speed with altitude, which is much less apparent in the 2017 data. The greatest temporal variation is for the highest altitudes (probed by HNC), revealing a dramatic ($\sim$ factor of two) wind speed reduction over the 9-month period between the two epochs.

{An additional feature of interest in the 2017 HNC Doppler map} is that the equatorial wind speed was slightly faster in the West ($-157\pm5$~\ms) than the East ($129\pm6$~\ms), which could indicate a slowing of the zonal jet as it traversed Titan's dayside-hemisphere.

\subsection{Deconvolved Wind Models}
\label{sec:models}

To further interpret these Doppler wind maps, it is necessary to account for (1) Titan's obliquity (26$^{\circ}$ toward the observer) and (2) beam convolution (smearing) due to the finite spatial resolution. We therefore performed a deconvolution/wind modeling procedure similar to \citet{lel19}, to derive an estimate for the intrinsic zonal wind velocities (${v}$) as a function of latitude ($\phi$), for each species. The ${v}(\phi)$ profiles were iteratively refined using Levenberg-Marquardt minimization to obtain the best fit to the data in Figure \ref{fig:obs}. After experimenting with different functional forms for the latitudinal wind speed profile, we decided on a Gaussian parameterization, which is able to reproduce the main features in our observed velocity maps, with a minimum number of free parameters: (1) the peak wind speed $v_{max}$, (2) the Gaussian FWHM, and (3) a latitudinal offset ($\phi_0$). Although more complex functional forms can be envisaged (allowing for longitudinal as well as additional latitudinal variability), the 1D Gaussian function provides a useful measure of the key wind field parameters of interest, without being subject to excessive bias due to noise, calibration or resolution limitations of our data.

A more detailed description of our wind profile retrieval procedure, along with a Figure showing the Best-fitting synthetic Doppler maps for each molecule, is given in Appendix B. Best-fitting, deconvolved ${v}(\phi)$ profiles are shown in the upper three panels of Figure \ref{fig:stability}.

\begin{table}
\centering
\caption{Best-fitting Gaussian wind model parameters \label{tab:pars}}
\begin{tabular}{lcccccc}
\hline\hline
Molecule & $\bar{z}$ (km) & Year & $L_s$ & $v_{max}$ (\ms) & $\phi_0$ ($^{\circ}$) & FWHM ($^{\circ}$) \\
\hline
CH$_3$CN &345& 2016 &$82^{\circ}$ &  $254\pm4$      & $-13\pm1$            & $88\pm3$\\ 
         && 2017 &$90^{\circ}$ & $185\pm3$      & $-11\pm1$           & $91\pm3$\\[1mm]  
HC$_3$N  &710& 2016 &$82^{\circ}$ &  $341\pm2$      & $-2\pm1$           & $91\pm1$\\
         && 2017 &$90^{\circ}$ & $225\pm2$      & $-1\pm1$           & $93\pm2$\\[1mm]
HNC      &990& 2016 &$82^{\circ}$ &  $373\pm6$      & $0\pm1$            & $70\pm2$\\
         && 2017 &$90^{\circ}$ & $196\pm2$      & $3\pm1$            & $101\pm2$\\ 
\hline
\end{tabular}
\end{table}

\section{Discussion}

At the upper-stratospheric/lower-mesospheric altitudes sounded by CH$_3$CN, the peak deconvolved zonal wind speed ($v_{max}$) dropped by $27\pm1$\% during the 9-month period spanned by the ALMA observations. A larger drop was observed for HC$_3$N ($34\pm1$\%) and larger-still for HNC ($47\pm1$\%). Thus, we see a significant trend for stronger temporal variability with increasing altitude above Titan's stratosphere. We also confirm the result of \citet{lel19}, that the lower-altitude zonal winds revealed by CH$_3$CN reach a peak velocity $11^{\circ}$--$13^{\circ}$ south of the equator (at both epochs).

Furthermore, while the shape of the CH$_3$CN and HC$_3$N wind profiles (characterized by the FWHM and $\phi_0$) did not undergo any significant changes from 2016 ($L_s=82^{\circ}$) to 2017 ($L_s=90^{\circ}$), the HNC wind profile was subject to substantial broadening, in addition to its dramatic slowing. This behaviour can be explained as a result of dynamical jet instabilities in Titan's upper atmosphere.

\begin{figure}
\begin{center}
\includegraphics[width=\textwidth]{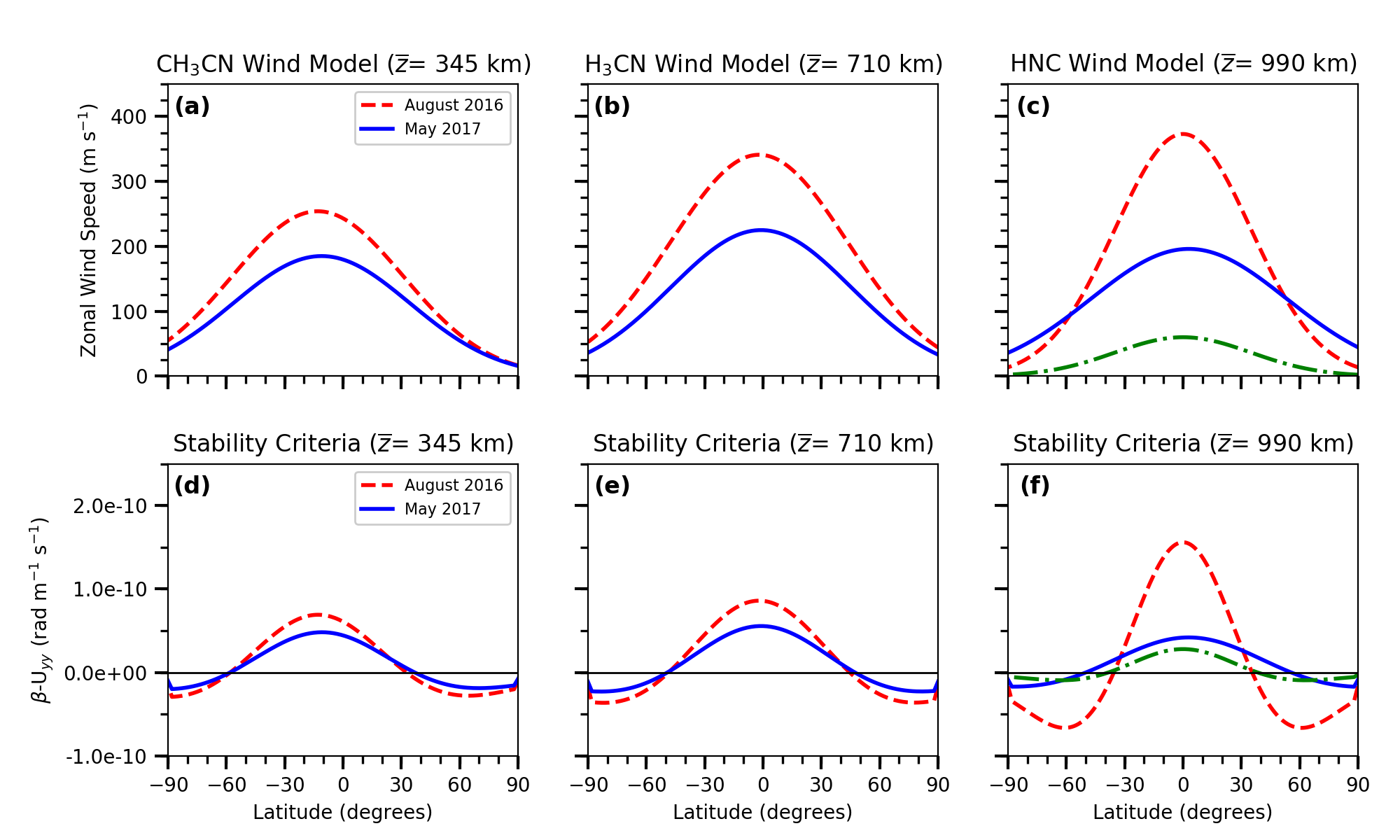}
\caption{Top row: deconvolved (retrieved) zonal wind speed profiles ($v(\phi)$) for each molecule, obtained in 2016 (red dashed) and 2017 (blue solid). The thermospheric model prediction of \citet{mul08} ($v_{max}=60$~\ms) is also shown (green dot-dashed curve), with Gaussian FWHM adopted from the 2016 HNC profile. Bottom row: corresponding jet instability analysis for each altitude and epoch, using the Rayleigh-Kuo criteria ($\beta-U_{yy}$). \label{fig:stability}}
\end{center}
\end{figure}

We investigated the HNC latitudinal wind profile (Figure \ref{fig:stability}, lower-right panel) for barotropic instability via the Rayleigh-Kuo criterion ($\beta-U_{yy}<0$), which compares the meridional gradient of the Coriolis force ($\beta$) to the second derivative of the zonal wind as a function of latitude ($U_{yy}$) \citep{1973_Kuo}. The meridional gradient in potential vorticity changes sign under similar conditions, which violates the Charney-Stern jet stability criteria \citep{1947_Charney, 1962_CharneyStern} and indicates that a jet is susceptible to baroclinic instability growth. The Rayleigh-Kuo criterion shows strong evidence for instability ($\beta-U_{yy}\ll0$) of the HNC thermospheric jet in 2016, as a result of its high speed and narrow FWHM. {Instability leads to a loss of momentum through the generation of eddies and waves at the jet flanks, causing the jet to slow and broaden towards the profile seen in 2017.  While ($\beta-U_{yy}<0$) still occurs at high latitudes in 2017, so further decay is possible, the slower, broader jets observed for all three species at the later epoch are expected to be less susceptible to the growth of instabilities.}

\begin{figure}
\begin{center}
\includegraphics[width=\textwidth]{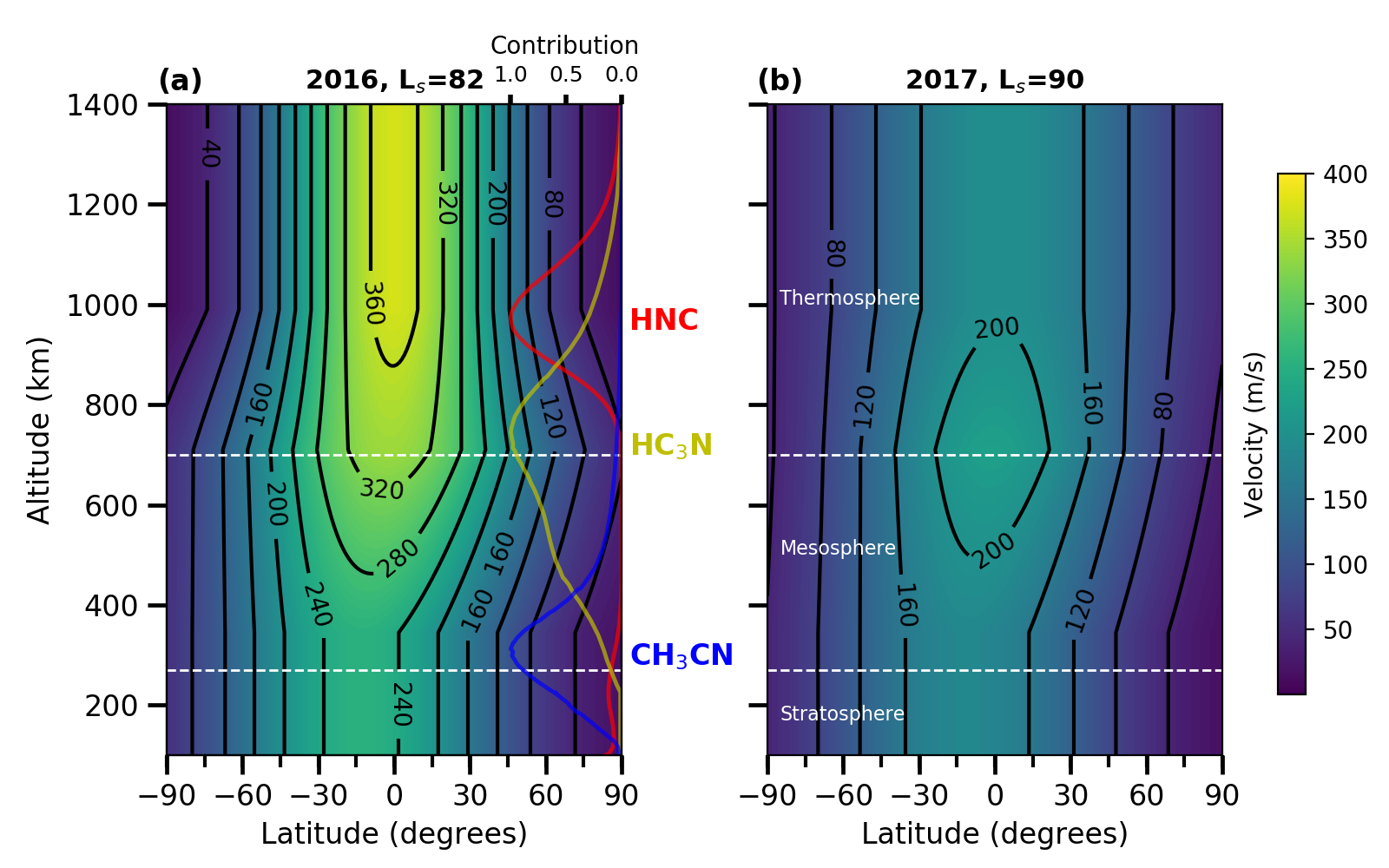}
\caption{Two-dimensional zonal wind fields derived from ALMA observations in 2016 ($L_s=82^{\circ}$) and 2017 ($L_s=90^{\circ}$). Peak-normalized emission contribution functions for each molecule (from \citet{lel19}) are shown for 2016 (upper-right x-axis) using colored curves. Wind speeds have been linearly interpolated between the peak contribution altitudes ($\bar{z}$) for the respective molecules, and are assumed constant outside the interpolation range (345--990~km). \label{fig:2dwinds}}
\end{center}
\end{figure}

Combining our derived $v(\phi)$ profiles with the peak contribution altitudes of \citet{lel19} (see Table \ref{tab:pars}), two-dimensional zonal wind fields were generated as a function of $\phi$ and $z$ (shown in Figure \ref{fig:2dwinds}). The fidelity of these maps is limited by (1) the coarse altitude sampling of our data, (2) smearing due to the broad range of altitudes contributing the the detected emission from each species, and (3) the assumption of Gaussianity in our $v(\phi)$ profiles. They nevertheless serve as a useful visualisation of our wind measurements.  The significant vertical wind shear (from thermosphere to stratosphere) observed in 2016 provides further evidence for baroclinic instability of Titan's high-altitude jets \citep{1947_Charney}.

It is only since the advent of ALMA that detailed, instantaneous global wind mapping of Titan has become possible. The previous ground-based mm-wave study by \citet{mor05} revealed equatorial zonal wind speeds of 60--160~\ms\ in Titan's stratosphere/mesosphere (decreasing in speed with higher altitudes). Despite the lower resolution and sensitivity of those data, a substantial increase in mesospheric wind speed is suggested during the 2003-2016 period (corresponding to $L_s=275^{\circ}$--$82^{\circ}$). It is interesting to note that the slower winds measured by \citet{mor05} were in the year following Titan's 2002 (northern winter) solstice ($L_s=270^{\circ}$); the slowing we observed in 2017 around the northern summer solstice ($L_s=90^{\circ}$) could therefore be related to a seasonally-recurring cycle in Titan's wind patterns and circulation, with fastest mesospheric/thermospheric wind speeds in the years leading up to the solstice. Due to the incomplete temporal sampling, however, the actual timescale of the thermospheric jet's variability remains unknown.

General circulation models (\emph{e.g.} \citet{new11,leb12}) produce stratospheric zonal winds in Titan's winter hemisphere as a consequence of temperature gradients at high winter latitudes. \citet{new11} showed that Titan's equatorial stratospheric superrotation is then maintained by repeated build-ups of instabilities in the low-latitude flank of this polar winter jet. These then trigger the generation of planetary Rossby waves that propagate from equatorial and summer low-latitudes through to the winter jet core, transporting negative angular momentum and thus causing an increase of angular momentum (spin-up) at low/equatorial latitudes. These `transfer events,' which last up to 16 Titan days ($\sim240$ Earth days), reduce the angular momentum of the polar jet, ultimately shutting down the source of instability, in a mechanism first described by \citet{1975_Gierasch, 1975_Rossow_Williams} (GRW). \citet{mul08} predicted thermospheric zonal wind speeds up to 60~\ms\ as a result of solar forcing, but this is insufficient to explain the ALMA observations. They also showed that thermospheric zonal winds may reach up to 120~\ms\ if the model is forced from below by (mesospheric) zonal winds of up to 50~\ms. The observed thermospheric jet variability could therefore arise as a consequence of intermittent GRW transfer events originating from lower altitudes. A complete general circulation model for Titan --- covering altitudes from the stratosphere to the thermosphere --- is needed in order to test this hypotheses.

Cassini CIRS observations of Titan's stratosphere and mesosphere revealed coldest south polar temperatures between 2012-2015 during formation of the winter polar vortex. Once the vortex is fully formed, atmospheric subsidence drives winter polar warming, first in the mesosphere (from early-2015) and later in the stratosphere (from mid-2016) \citep{tea17,tea19}. The associated mesospheric zonal winds were strongest from mid-2012 to late-2016, and weakened thereafter \citep{sha20,vin20}, {consistent with our ALMA observations}. We speculate that the strong, dynamically-unstable thermospheric jet seen by ALMA in 2016 could have been driven by gravity waves originating from high wind shear and rapid changes in the middle-atmospheric temperature and circulation. The observed reduction in middle-atmosphere wind shear during 2016-2017 \citep{sha20} could then have contributed to a reduction in thermospheric jet speeds \emph{via} a reduction in wave breaking once the polar vortex was fully formed and in a stable state.

{Titan's stratospheric (winter hemisphere) jet/vortex is analogous to Earth's polar vortex --- another potential source of gravity waves \citep{yos00,sat08} that could influence the upper-atmospheric circulation. The discovery of rapid dynamical variability in Titan's upper atmosphere may therefore be important for the development of theories linking the middle and upper-atmospheric circulation on similar (Earth-like) planets throughout the Galaxy.}

\section{Conclusion}

Capitalizing on the unique sensitivity and resolution of ALMA, we have generated Doppler maps of three different molecules, from which Titan's zonal wind field was derived as a function of latitude and altitude,  in the range $z\sim350$--$1000$~km (upper stratosphere to thermosphere). Rapid changes in wind speeds were observed over a 9-month period, the most striking of which is a 47\% drop in the speed of Titan's thermospheric jet, accompanied by a loss of latitudinal confinement. This {may be} explained as a result of dynamical instabilities in the thermosphere in 2016, combined with a seasonal/intermittent loss of forcing by gravity waves from the middle atmosphere. {Such strong time-variability is unexpected and provides a new challenge for our understanding of upper-atmospheric dynamics in terrestrial-like atmospheres}. Additional, temporally and spatially resolved wind observations over the coming years will be vital to elucidate the magnitude and cadence of variations in these high-altitude winds. This will provide crucial input for general circulation models, enabling us to test the hypothesis that Titan's transient thermospheric jet can be driven by vertical and latitudinal transport of angular momentum via gravity waves.

\acknowledgments

This work was supported by NSF Grant AST-1613987, NASA's Solar System Observations program, the NASA Astrobiology Institute, the UK Science and Technology Facilities Council and NASA OPR grant \#80NSSC18K0108 (JPL subcontract 1643481). It makes use of ALMA data sets ADS/JAO.ALMA\#2015.1.01023.S and \#2016.A.00014.S. ALMA is a partnership of ESO, NSF (USA), NINS (Japan), NRC (Canada), NSC and ASIAA (Taiwan) and KASI (Republic of Korea), in cooperation with the Republic of Chile. The JAO is operated by ESO, AUI/NRAO and NAOJ.  The NRAO is a facility of the National Science Foundation operated under cooperative agreement by Associated Universities, Inc.

%

\vspace{5mm}
\facilities{ALMA}


\software{atmWindMapGaussFit \citep{zen20c}}



\appendix

\section{Doppler Velocity Error Maps}

\begin{figure}[h!]
\includegraphics[height=50mm]{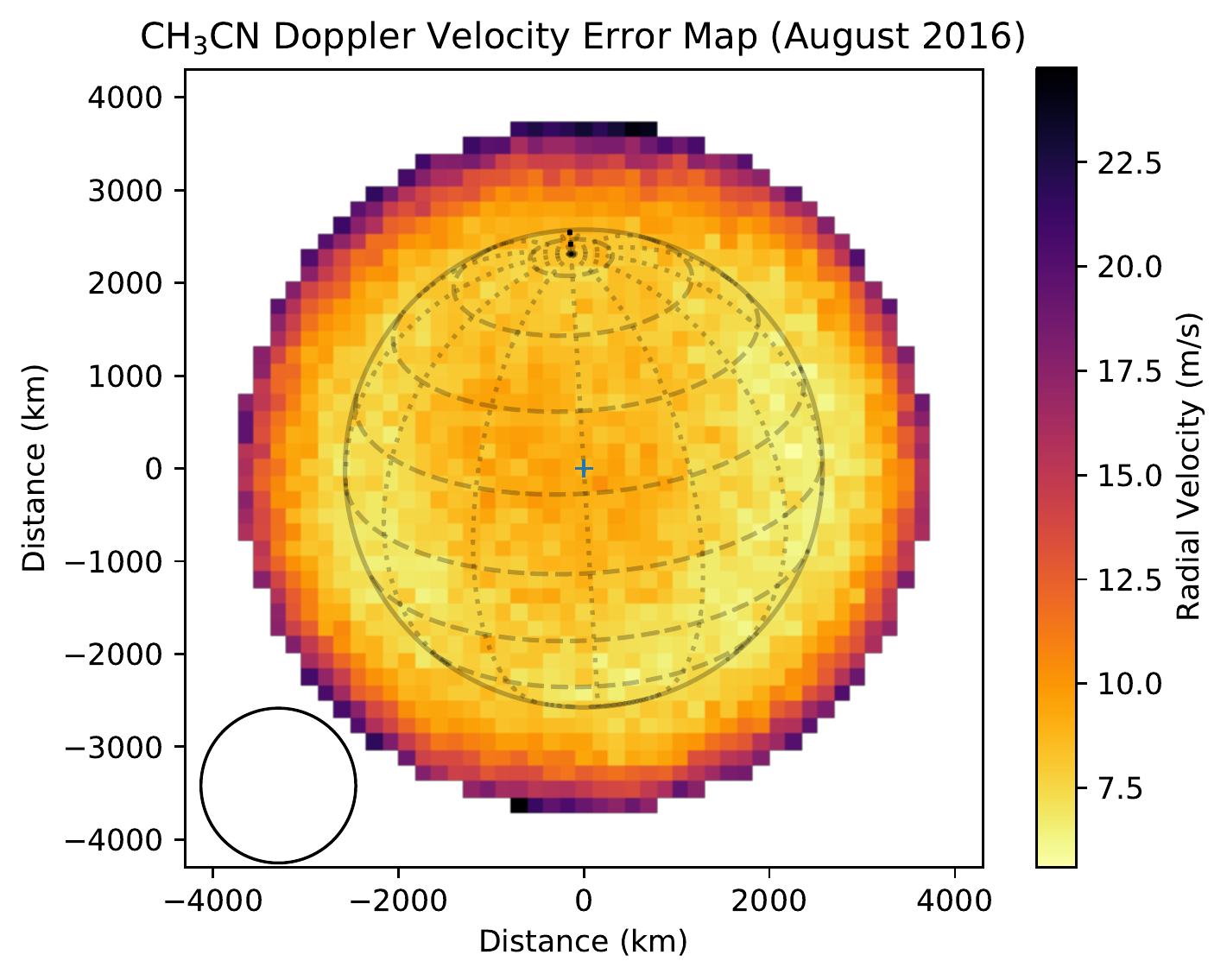}
\includegraphics[height=50mm]{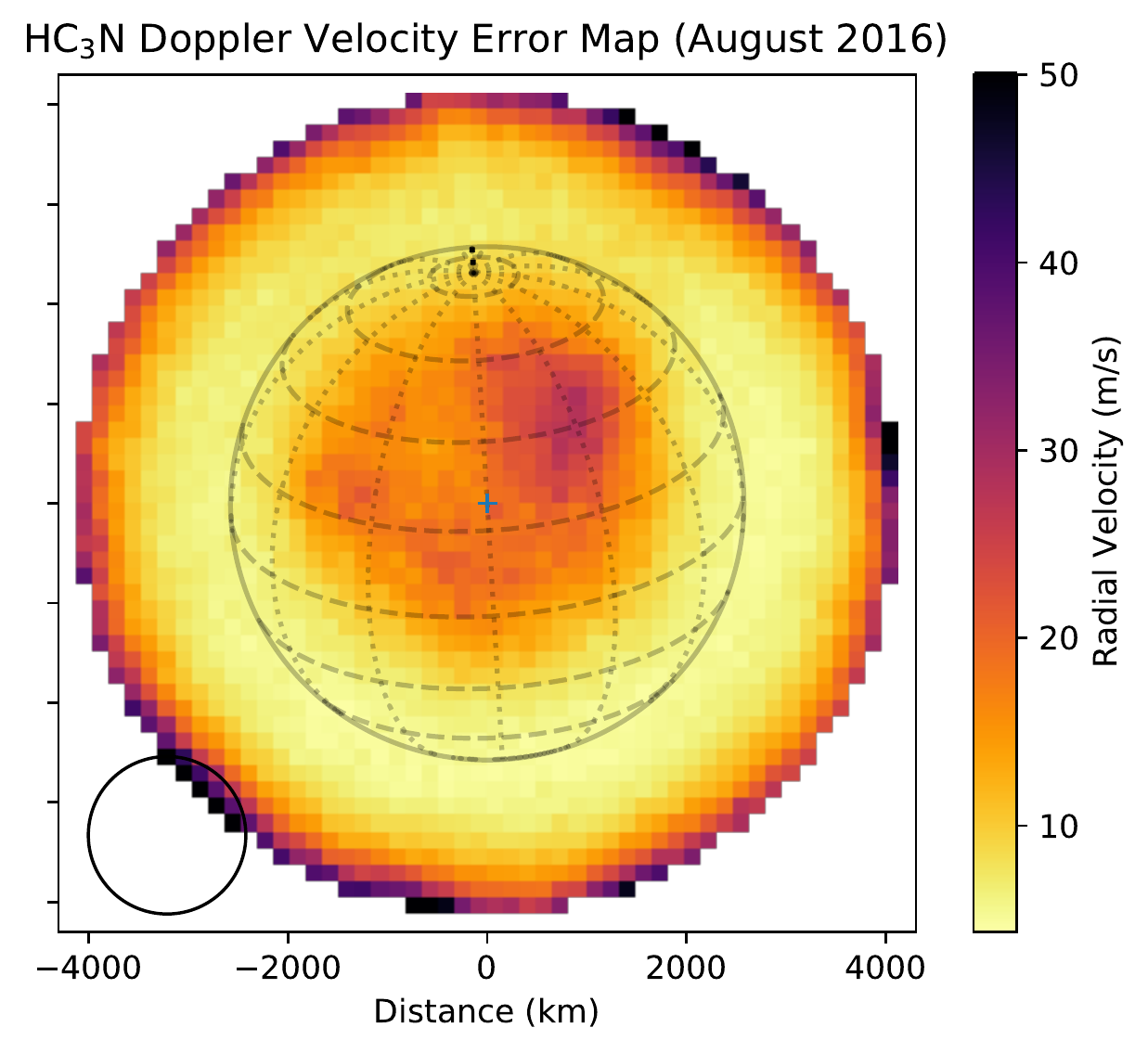}
\includegraphics[height=50mm]{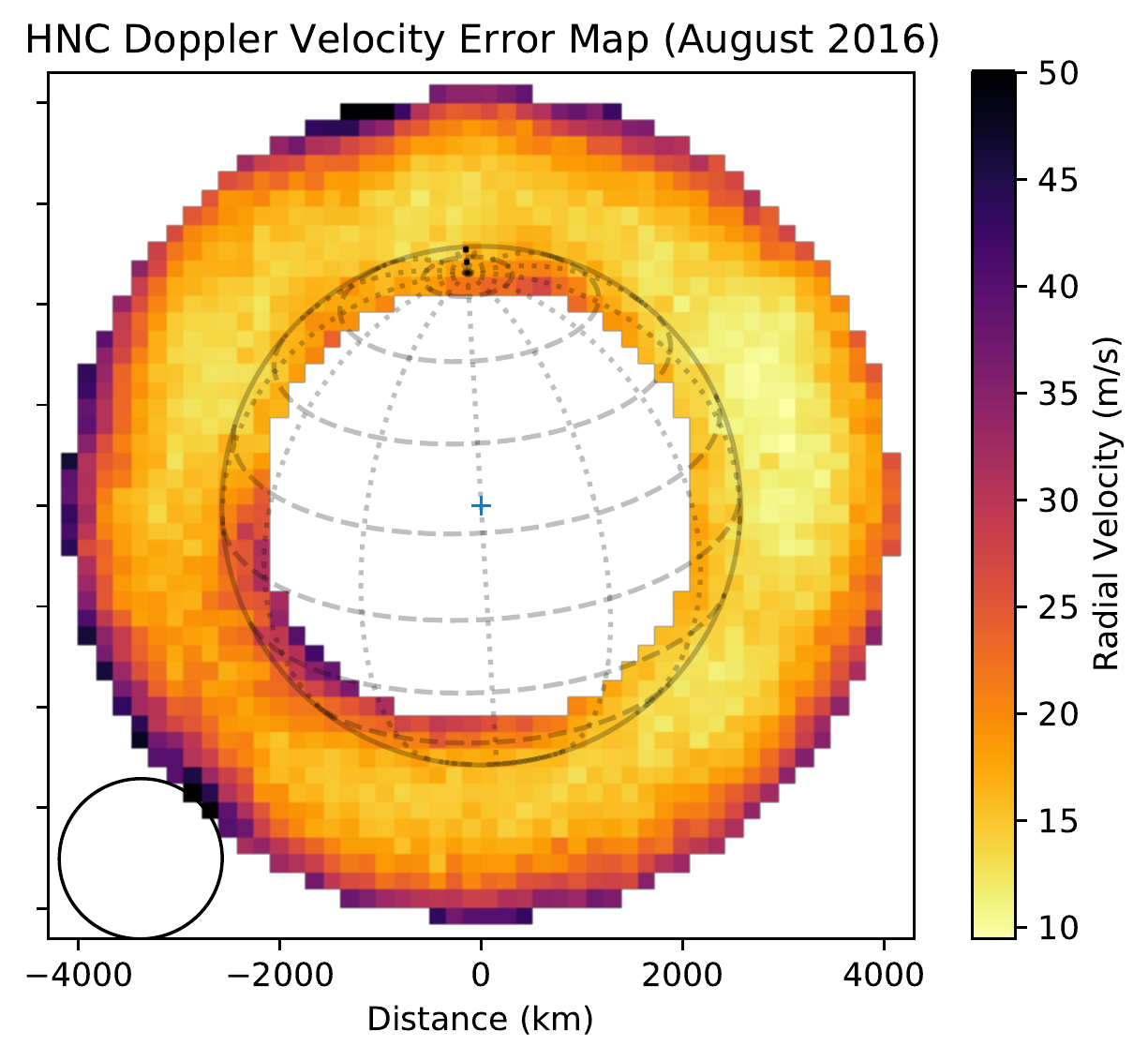}\\
\includegraphics[height=50mm]{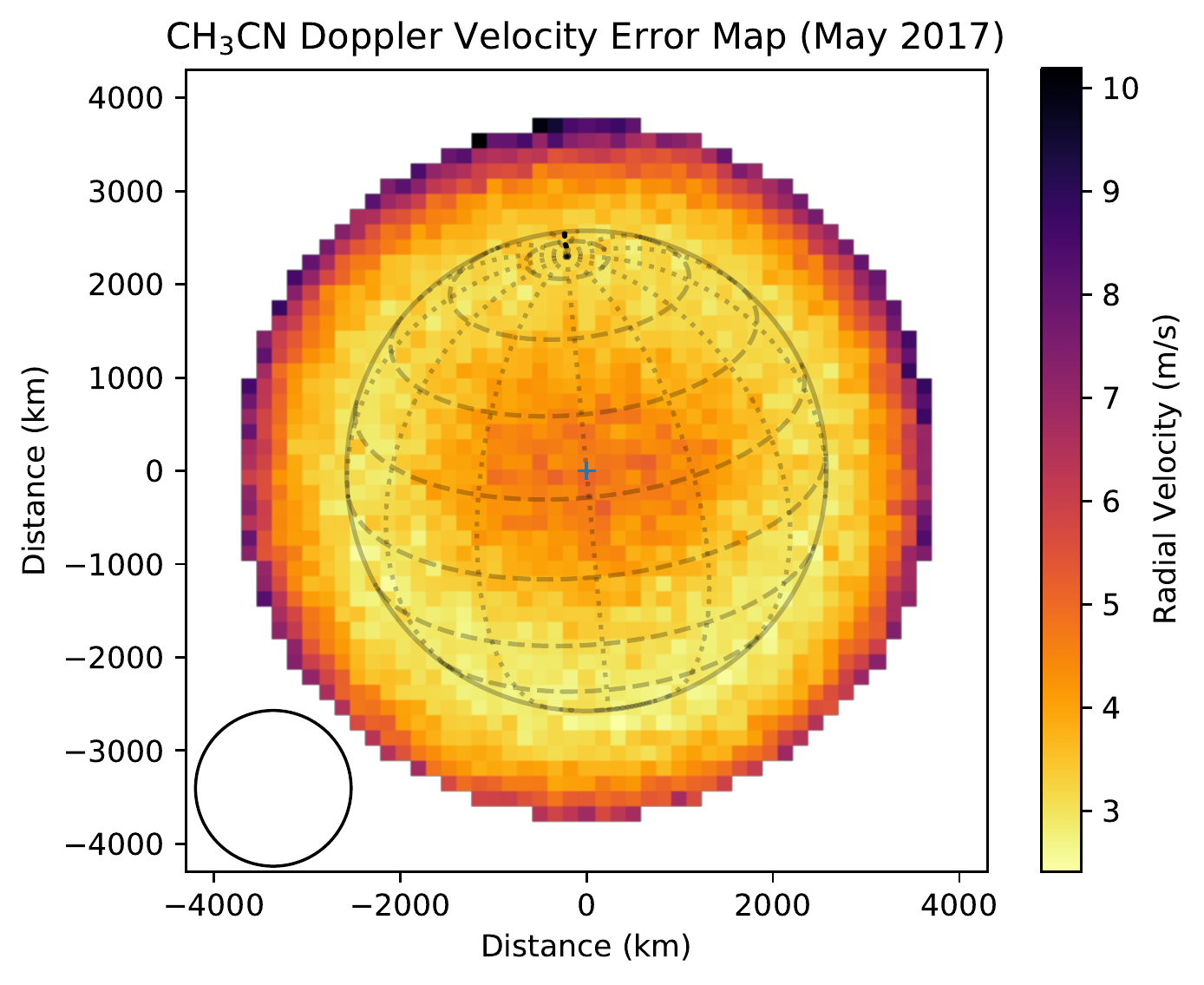}
\hspace*{1mm}
\includegraphics[height=50mm]{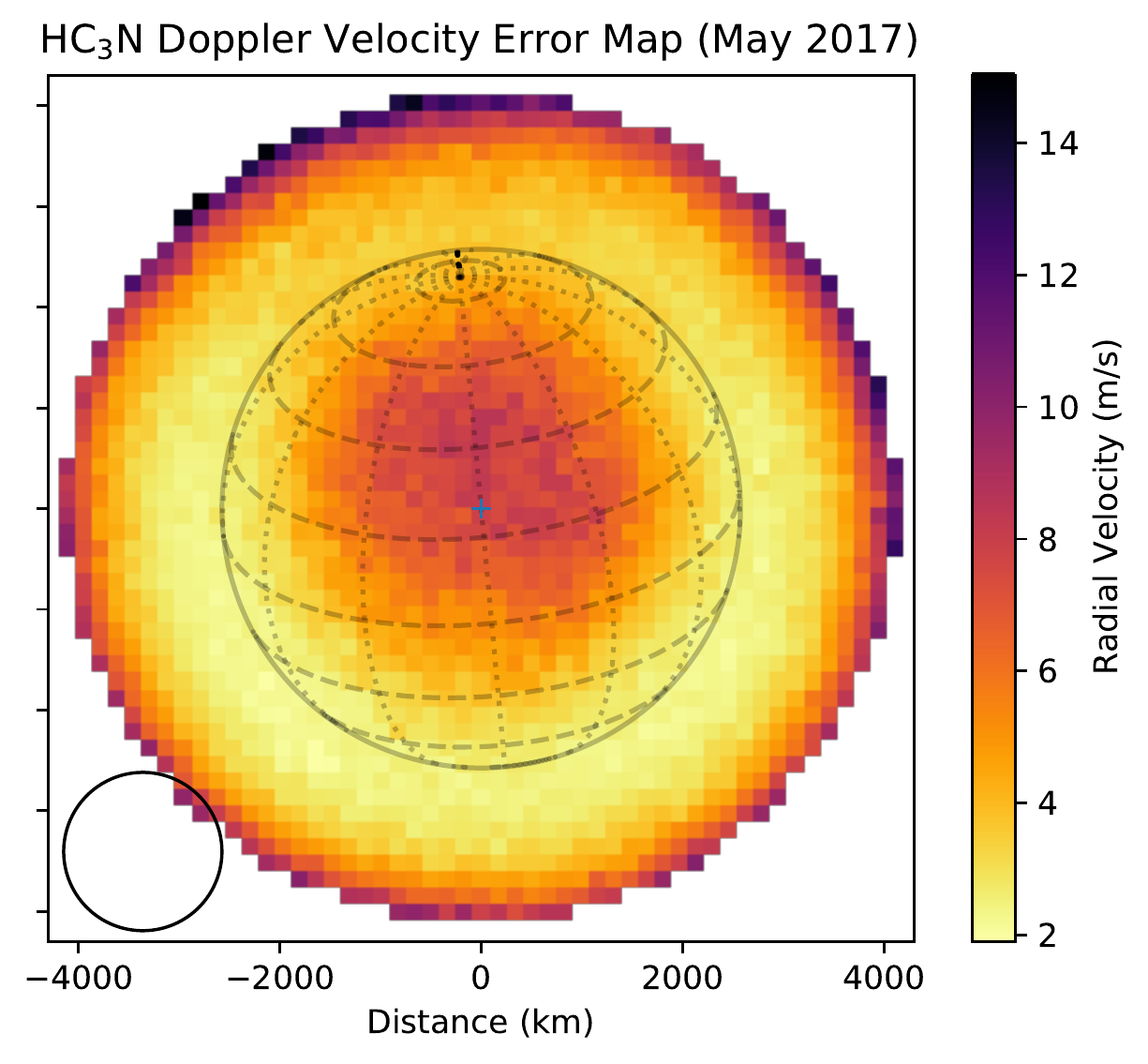}
\includegraphics[height=50mm]{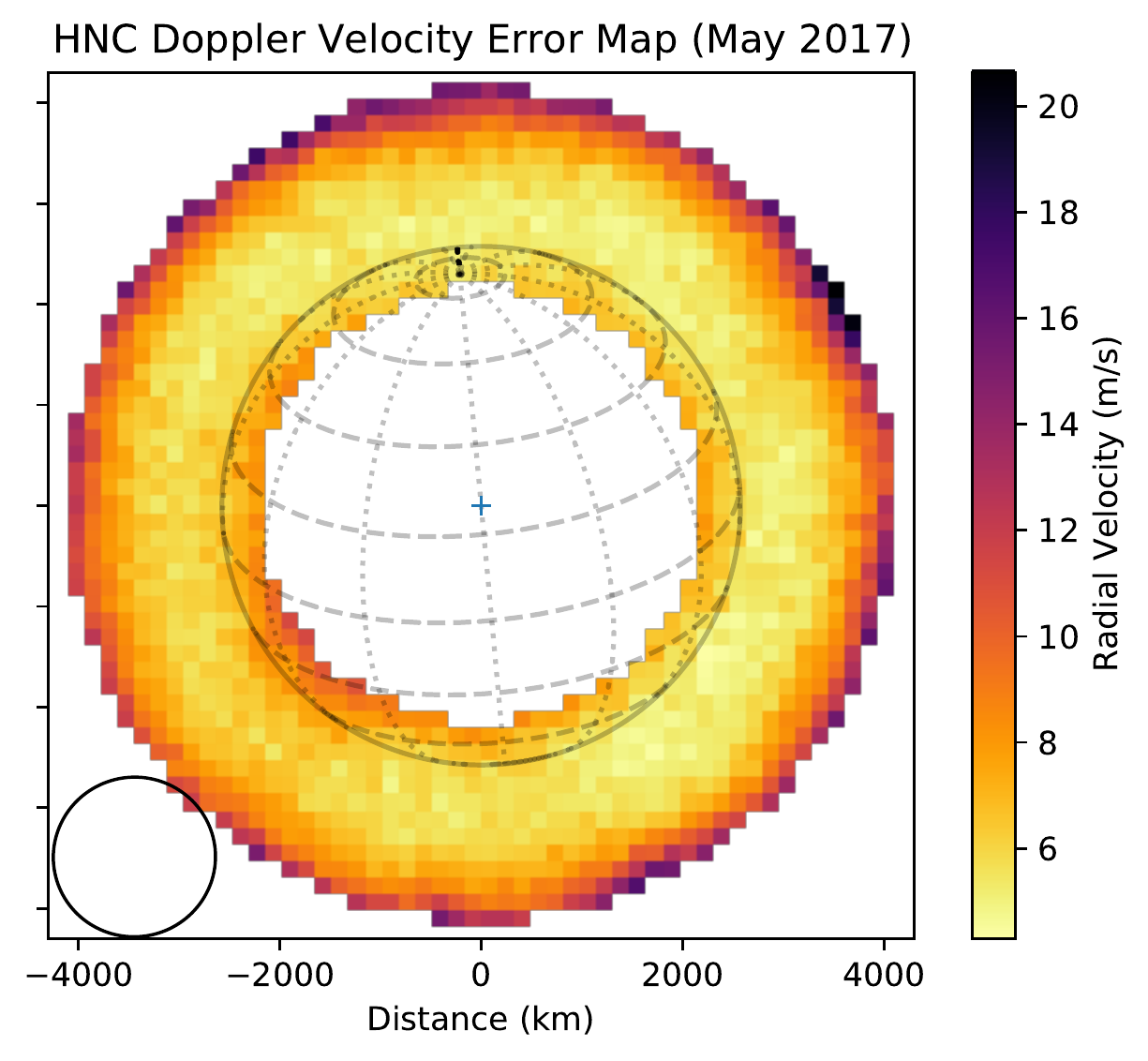}
\caption{Doppler velocity error maps for each molecule in 2016 and 2017. These are the 16-84th percentile ranges of the distribution of velocities obtained from 300 Monte Carlo error runs per pixel (corresponding to $1\sigma$ velocity uncertainties, assuming Gaussian statistics). \label{fig:errormaps}}
\end{figure}

{Line-of-sight Doppler velocities were derived for each molecule on both epochs using Gaussian fits to the spectra extracted from each spatial pixel in the ALMA images. Velocity uncertainties were derived using a Monte Carlo approach (Section 2.1), and the resulting error maps are shown in Figure \ref{fig:errormaps}. }

\section{Wind Velocity Retrievals}

The Doppler wind velocity maps in Figure 2 of the main text were deconvolved to obtain estimates for Titan's intrinsic (underlying) zonal wind field ${v}(\phi)$ for each molecule. Our method is based on \citet{lel19} and involved construction of a 3D grid of (100~km wide) cubic cells, covering Titan's entire atmosphere. Each cell was assigned a zonal wind velocity vector ($\vec{v}(\phi)$), an abundance, and a temperature (the latter two quantities were obtained from the limb retrievals of \citet{cor19}). {The derived ${v}(\phi)$ profile for each molecule is assumed not to vary with height, and therefore represents an average over the altitudes contributing to the emission from that molecule. This is reasonable given that the wind contribution functions are quite strongly peaked \citep{lel19}, so the emission is localized from a limited altitude range, over which the velocity is not expected to vary strongly.  Nevertheless, the possibility that the retrieved velocity profiles could be biased by emission from altitudes different from the mean wind contribution function altitude should be considered when interpreting the results}.  

The spectrum from each image pixel (projected in 2D in the plane of the sky) was calculated by integrating the equation of radiative transfer \citep{van07} along the line of sight, and the resulting image cube was convolved and resampled to the appropriate beam size (and pixel scale) for the observed molecular transitions. Finally, the convolved image cube was integrated along the spectral axis to determine the centroid velocity of each pixel. The sum of squares of the residuals (model minus observations) was minimized using the MPFIT algorithm \citep{mar12} to determine the optimal set of wind profile parameters [$v_{max}$, FWHM and $\phi_0$] (assuming a Gaussian ${v}(\phi)$ profile), which are given for each molecule and epoch in Table \ref{tab:pars}. The ($1\sigma$) uncertainties on each parameter were obtained from the MPFIT covariance matrix. The associated best-fitting ${v}(\phi)$ profiles are shown in the upper three panels of Figure \ref{fig:stability}, and corresponding synthetic Doppler wind maps are shown in Figure 6. A good match to the observations (Figure 2) is evident in each case, {\bf and for 2016, our synthetic wind maps compare well with those in Supplementary Fig. 7 of \citet{lel19}}. 

The code used to perform these wind speed profile retrievals is available for download from zenodo.org \citep{zen20c}.

\begin{figure}[h!]
\includegraphics[height=50mm]{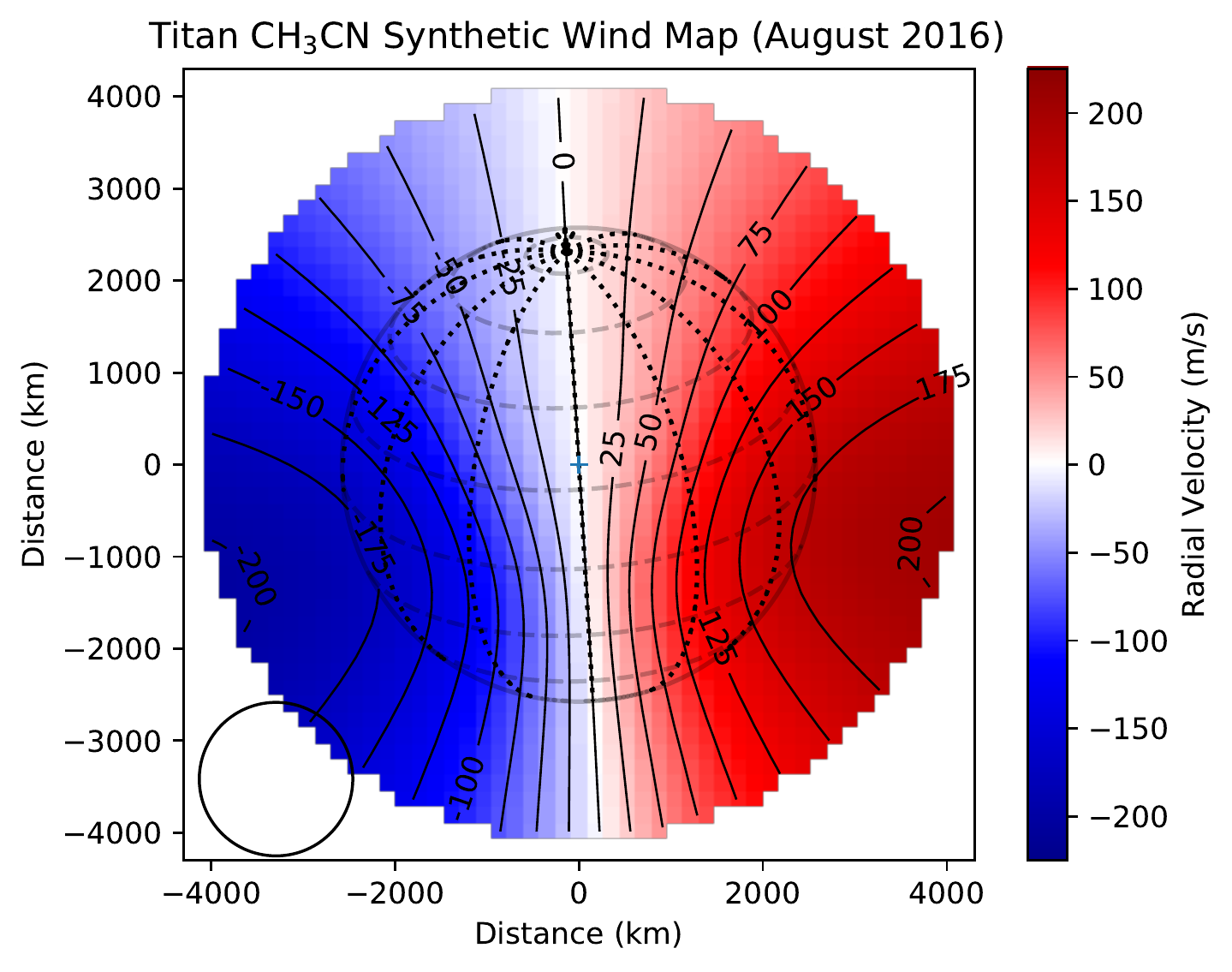}
\includegraphics[height=50mm]{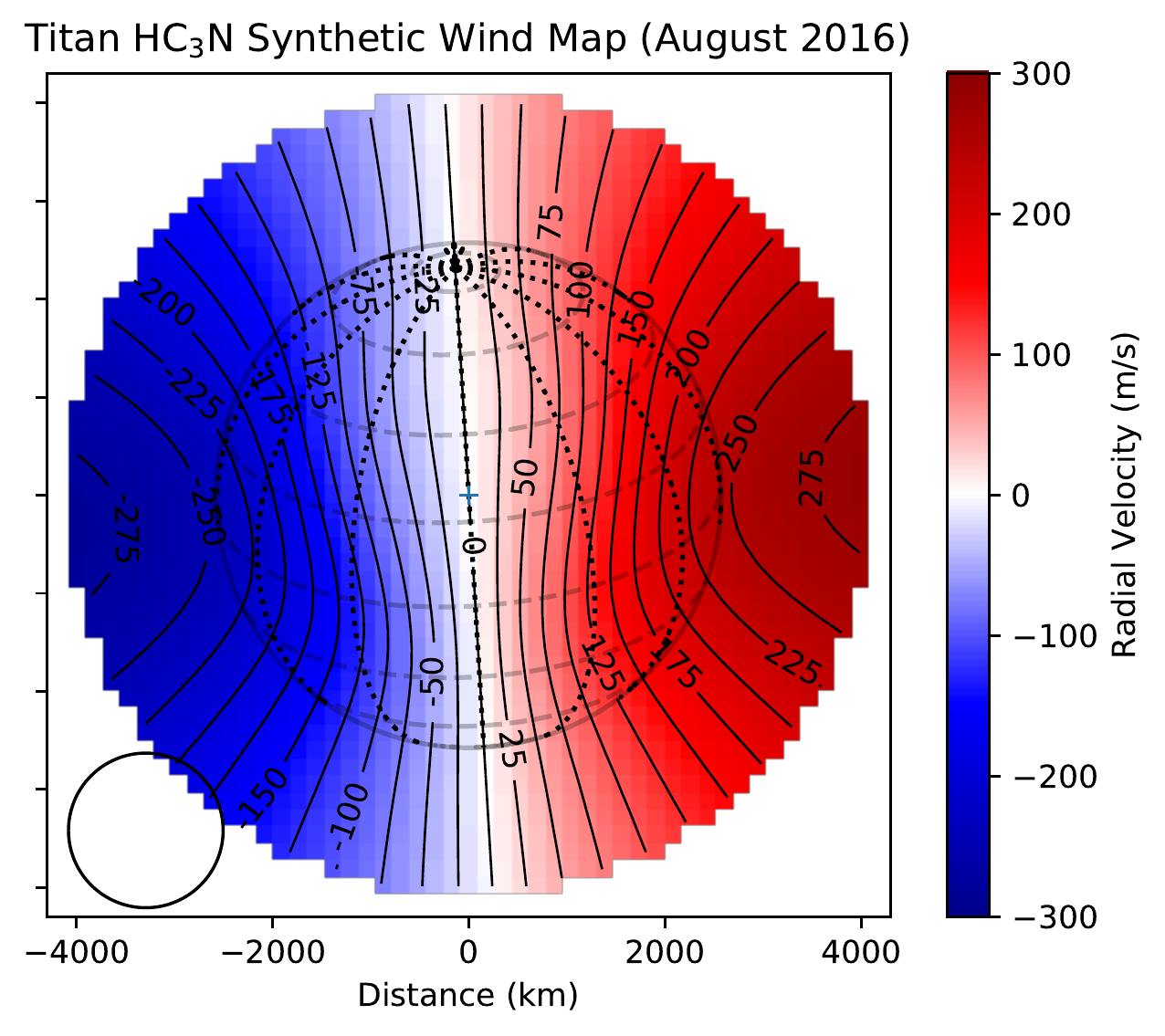}
\includegraphics[height=50mm]{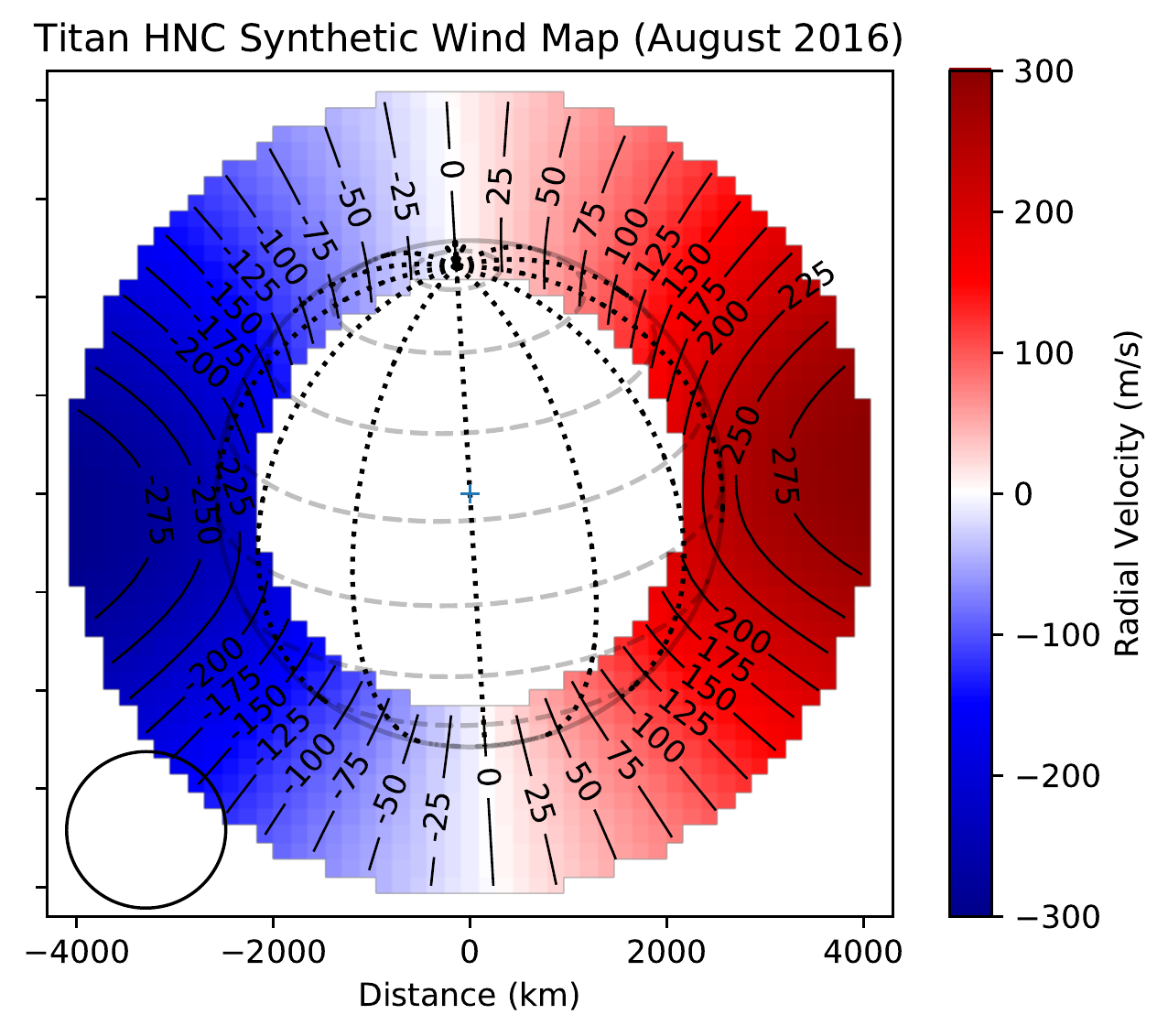}\\
\includegraphics[height=50mm]{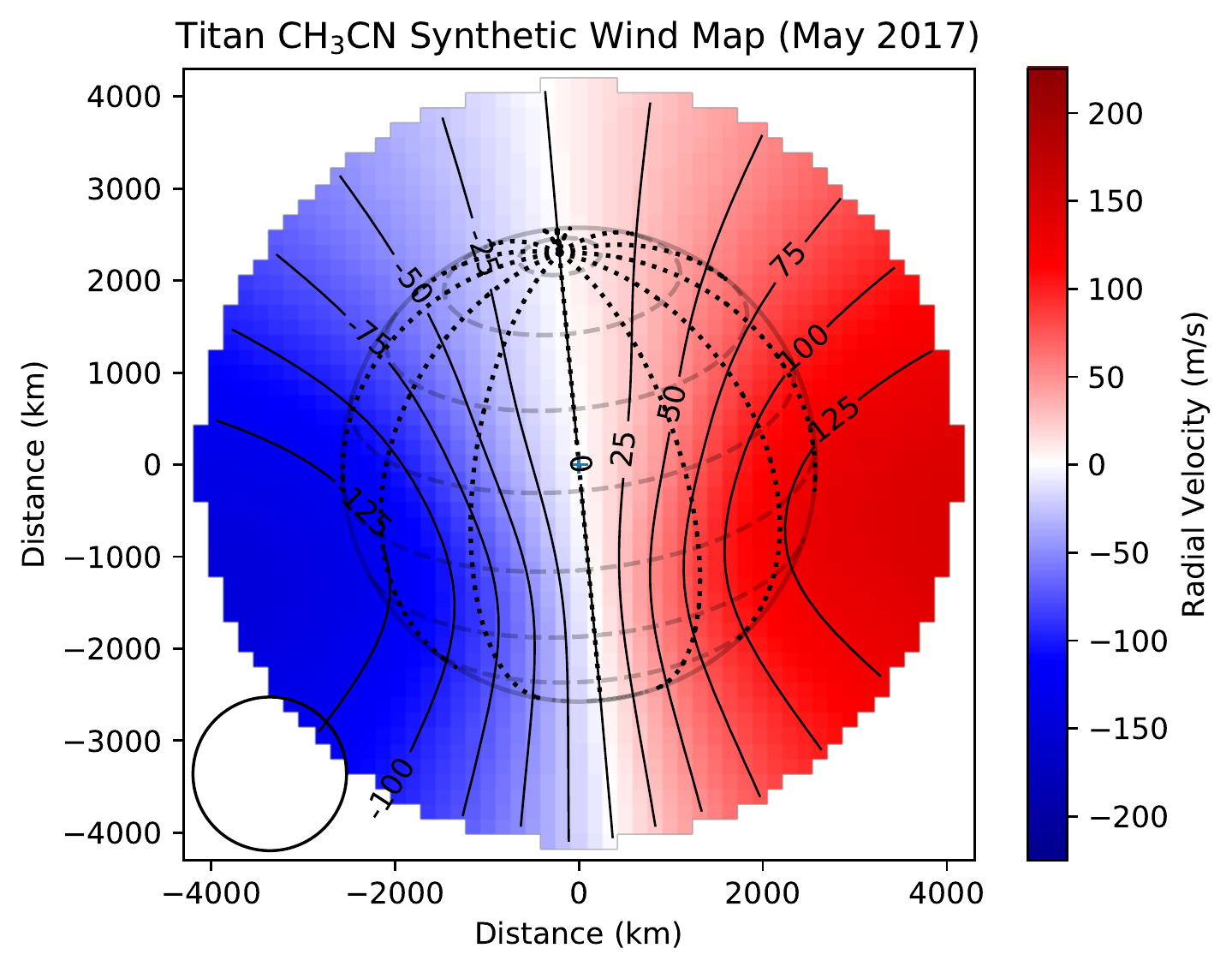}
\includegraphics[height=50mm]{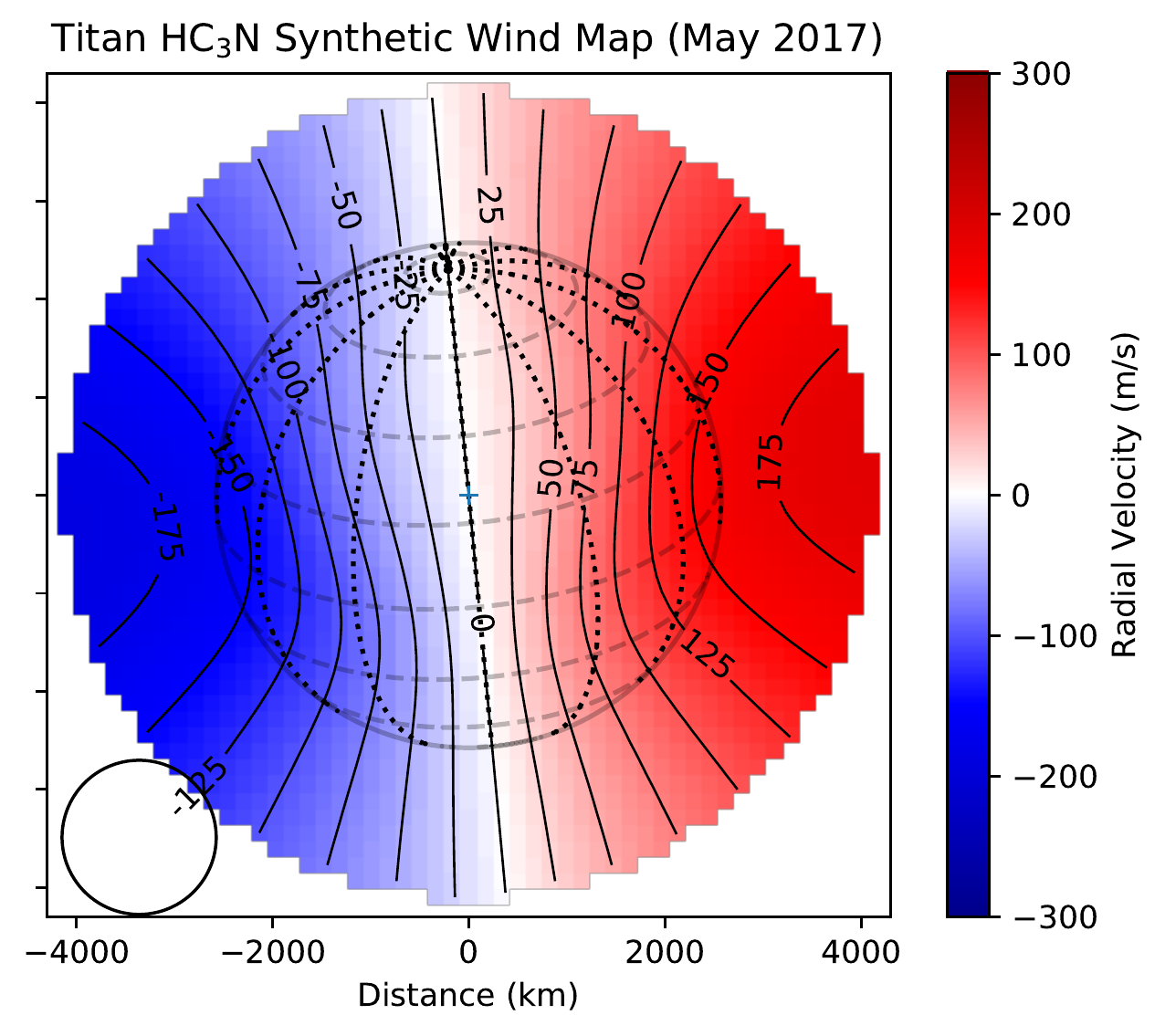}
\includegraphics[height=50mm]{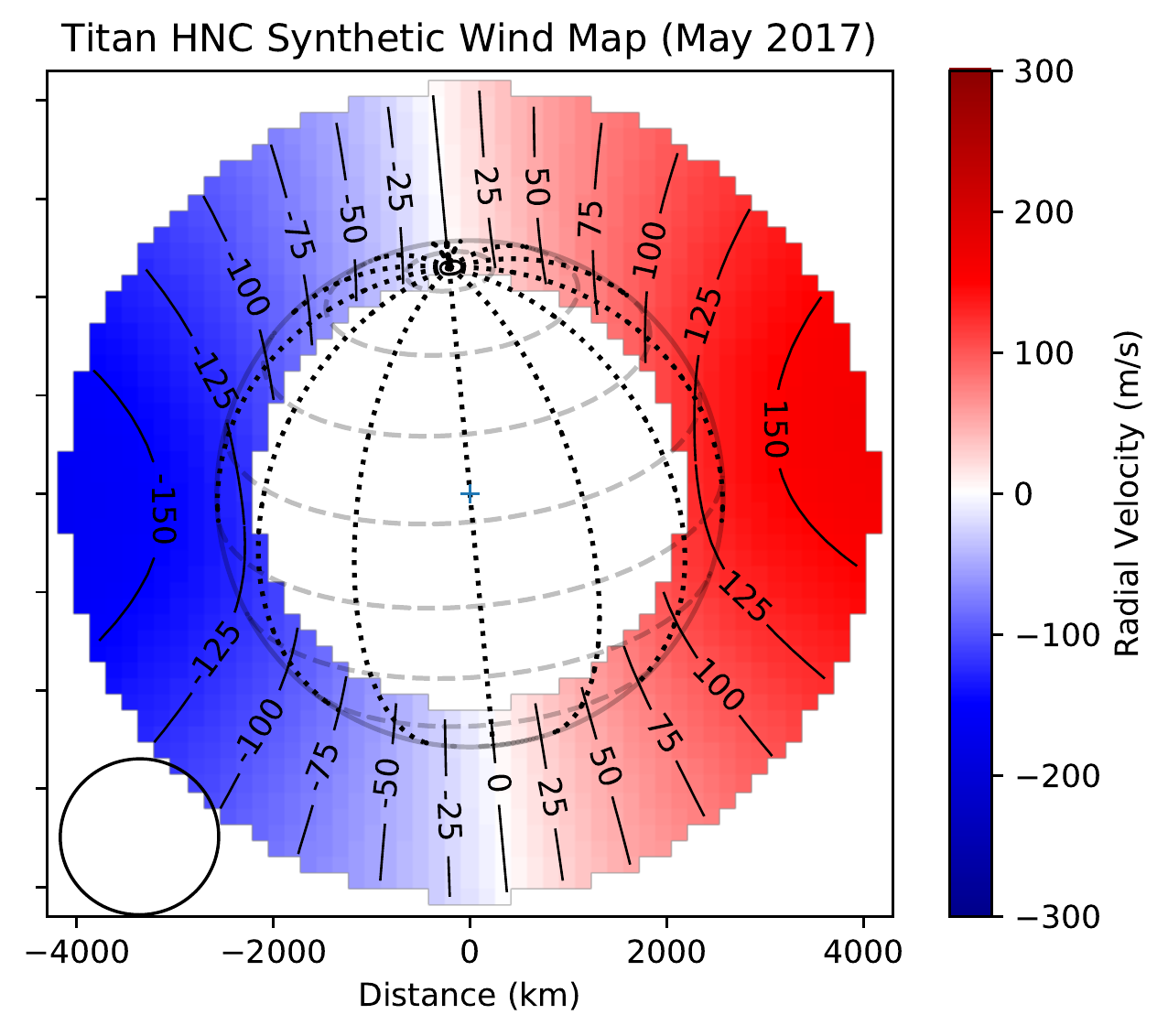}
\caption{Best-fitting synthetic (convolved) wind fields derived for each molecule in 2016 and 2017, assuming a Gaussian zonal wind velocity profile as a function of latitude. Velocity contours are labeled in units of m\,s$^{-1}$. Wire-frame sphere indicates Titan's orientation with respect to the ALMA field of view. \label{fig:models}}
\end{figure}

\bibliography{TitanRefs}{}
\bibliographystyle{aasjournal}



\end{document}